\documentclass[article]{JHEP}
\usepackage{amsmath,epsfig}
\usepackage{amssymb,amsfonts}
\usepackage{epsfig}
\relax
\renewcommand{\theequation}{\arabic{section}.\arabic{subsection}.\arabic{equation}}
\def\be{\begin{equation}}
\def\ee{\end{equation}}
\def\bea{\begin{eqnarray}}
\def\eea{\end{eqnarray}}
\def\be{\begin{equation}}
\def\ee{\end{equation}}

\newcommand\fverb{\setbox\pippobox=\hbox\bgroup\verb}
\newcommand\fverbdo{\egroup\medskip\noindent%
                        \fbox{\unhbox\pippobox}\ }
\newcommand\fverbit{\egroup\item[\fbox{\unhbox\pippobox}]}
\newbox\pippobox
\def\F{\Phi}

\def\h{\eta}
\def\ka{\kappa}
\def\z{\zeta}
\def\m{\mu}
\def\n{\nu}
\def\r{\rho}
\def\s{\sigma}
\def\t{\theta}

\def\p{\partial}

\def\f{\varphi}
\def\a{\alpha}
\def\b{\beta}
\def\d{\delta}

\def\g{\gamma}
\def\G{\Gamma}

\def\ba{\begin{eqnarray}}
\def\ea{\end{eqnarray}}
\def\nb{\nonumber}
\def\jh{{\hat{\jmath}}}

\def\ie{{\it i.e.}}

\newcommand{\beq}{\begin{equation}}
\newcommand{\eeq}{\end{equation}}

\newcommand{\bx}{\bar{x}}
\newcommand{\bz}{\bar{z}}
\newcommand{\by}{\bar{y}}

\newcommand{\tl}{\tilde{l}}

\newcommand{\pd}{\partial}
\newcommand{\beqa}{\begin{eqnarray}}
\newcommand{\eeqa}{\end{eqnarray}}
\newcommand{\hj}{\hat{\jmath}}

\title{String
amplitudes in the Hpp-wave limit of $AdS_3\times S^3$}
\author{ M. Bianchi$\ ^1$, G. D'Appollonio$\ ^2$,  E.
Kiritsis$\ ^{3,4}$, O. Zapata$\ ^1$\\
$~$\\
$~$\\
$^1~$Dipartimento di Fisica, \ Universit{\`a} di Roma \ ``Tor
Vergata''
\\
I.N.F.N.\ -\ Sezione di Roma \ ``Tor Vergata''
\\
Via della Ricerca  Scientifica, 1 - 00133 \ Roma, \ ITALY
\\
$~$ \\
$^2~$ LPTHE, Universit\'e Paris VI, 4 pl Jussieu,
\\
75252 Paris cedex 05, FRANCE\\
$~$\\
$^3~$CPHT, UMR du CNRS 7644, Ecole Polytechnique,\\
91128, Palaiseau, FRANCE\\
$~$\\
$^4~$Department of Physics, University of Crete\\
71003 Heraklion, GREECE\\
$~$\\
{\tt E-mails:
massimo.bianchi@roma2.infn.it, giuseppe@lpthe.jussieu.fr,
kiritsis@cpht.polytechnique.fr, zapata@roma2.infn.it} }

\preprint{\hepth{0402004} \\ CPHT-RR117.1203  \\ NSF-KITP-04-05
\\ LPTHE-03-38 \\ ROM2F/2003/37}

\abstract{We compute  string amplitudes on pp-waves supported by
NS-NS 3-form fluxes and arising in the Penrose limit of
$AdS_3\times S^3\times {\mathcal M}$. We clarify the role of the
non-chiral accidental $SU(2)$ symmetry of the background. We
 comment on the extension of our results to the superstring
and propose a holographic formula in the BMN limit of the
$AdS_3/CFT_2$ correspondence valid for any correlator.}

\begin{document}

\maketitle 

\section{Introduction}

The $AdS/CFT$ correspondence suggests a holographic duality
between (super)string theory on anti-de Sitter spaces (AdS) and
(super)conformal field theories defined on their boundaries. In
the past few years, Maldacena's original conjecture has passed a
large number of tests, especially in the $AdS_5/CFT_4$ case, and
has been a precious source of insights on field theories in their
(super)conformal phase, along RG flows where (super) conformal
symmetry is broken at tree level either spontaneously or explicitly, and
to some extent even in cases with a dynamically generated scale (see
\cite{agmoo,dzfdh} and
reference therein).

The main obstacle towards extending the holographic duality beyond
the (super)gravity approximation that captures the strong coupling
regime of the boundary (conformal) field theory is represented by our
limited understanding of how to quantize the superstring in the
presence of R-R backgrounds \cite{berkovits}. One possible
exception is the background $AdS_3\times S^3\times
\mathcal{M}$ supported by a NS-NS 3-form flux which is the near
horizon geometry of a bound-state of fundamental strings (F1) and
penta-branes (NS5) \cite{gks}. Powerful CFT techniques can be
exploited in this case to compute the spectrum and string amplitudes since the
dynamics on the world-sheet is governed by an $SL(2,\mathbb{R})\times SU(2)$
WZNW model \cite{mo1,mojs2,mo3}. S-duality relates NS-NS 3-form
flux to R-R flux or a combination of the two and one may in
principle resort to the hybrid formalism of Berkovits, Vafa and Witten
to make part of the space-time supersymmetries manifest \cite{bvw} and
compute some three-point amplitudes \cite{bobdol}.

The dual two-dimensional superconformal field theory is expected
to be  the non-linear $\sigma$-model with target space the
symmetric orbifold ${\mathcal M}^N/{\mathcal S}_N$
\cite{malda,agmoo,gks}, where $\mathcal M$ is taken to be either $T^4$
or $K3$. Quantitative comparison with boundary conformal field
theory predictions is hampered by the presence of non-compact
directions in the target space of the non-linear $\sigma$-model
\cite{mo3,sw} which would be lifted by turning on a R-R
background, {\it i.e.} moving away from the symmetric orbifold
point in the moduli space.

An interesting limit of $AdS\times S$ is the plane wave (pp-wave),
which corresponds to the local background seen by an observer
moving at the speed of light in the original space. This procedure
of zooming-in around a null geodesic is known as Penrose limit
\cite{ppwaves}. Remarkably many pp-waves are amenable to
quantization in the light-cone gauge even in the presence of R-R
backgrounds \cite{metsatseyt}. The string spectrum for $p^+\neq 0$
can be computed exactly and contrasted with the spectrum of
operators with large R-charge $J$ that survive the so-called BMN limit
(large $N$ and $J$ with $J^2/N$ fixed). Unfortunately, string
interactions and the spectrum at $p^+= 0$ are hard to determine
in the light-cone gauge \cite{ppreviews}.

Once again it is fruitful to consider the Hpp-wave resulting from
the
Penrose limit of $AdS_3\times S^3$ supported by a NS-NS 3-form
flux. The world-sheet CFT for the bosonic coordinates is a
six-dimensional generalization \cite{kehag} of the Nappi-Witten
(NW) model \cite{nw}. The relevant Heisenberg current algebra is
$\widehat{\mathcal H}_6$, or actually two copies of
$\widehat{\mathcal H}_6$ with a common central element and a
non-chiral external $SU(2)$ automorphism preserved by the limiting
flux.  This is broken to $U(1)$ when the NS-NS fluxes through the two planes
are different,
$H_{+12} = \mu_1 \neq
H_{+34}=\mu_2$. The theory depends on $\mu_1/\mu_2$ only and
is  exactly solvable for all values of $\mu_1,\mu_2$.
{}From the current algebra point of view, the
Penrose limit is carried out by contracting the currents of
$\widehat{SL}(2,\mathbb{R})_{k_1}\times \widehat{SU}(2)_{k_2}$ with
$\mu_1^2 k_1 = \mu_2^2 k_2$. This marginal deformation
interpolates between the generic 6-d Hpp-wave ($\mu_1 \neq
\mu_2$), the (super)symmetric one ($\mu_1 = \mu_2$) and the NW
model ($\mu_1 = 0$ or $\mu_2=0$) or even flat space-time ($\mu_1 =
0$ and $\mu_2=0$) very much as the `null deformation' discussed in
\cite{ikp} interpolates between $AdS_3\times S^3$ and $R^+\times
S^3$ with a linear dilaton before any Penrose limit is taken.

Exploiting  current algebra techniques and a quasi-free field
resolution \cite{kk}, it was possible \cite{dak}  to explicitly
compute string amplitudes in the closely related NW model, that
represents the Penrose limit of the near-horizon geometry of a
stack of NS5-branes \cite{gomis,kirpio} and realizes the
$\widehat{\mathcal H}_4$ current algebra.

In the present paper, we will apply the same techniques to the
pp-wave representing the Penrose limit of $AdS_3\times S^3\times
{\mathcal M}$. Although we will almost exclusively concentrate our
attention on the bosonic string, we will briefly comment on how to
extend our results to the superstring. We will
compute two-, three- and four-point amplitudes with insertions of
tachyon vertex operators of any of the three types of representations
of the $\widehat{\mathcal H}_6$ current algebra: actually
depending on the value of the light-cone momentum $p^+$, the states
belong to discrete representations when $p^+\neq 0 $ or to continuous
representations when $p^+=0$.

The main novelties we found with respect to the
$\widehat{\mathcal H}_4$ case are the existence
of non-chiral symmetries which correspond to background isometries
not realized by the zero-modes of the currents
and the presence in the spectrum of new representations
of the current algebra that satisfy a modified highest weight condition.
The results are compactly encoded in terms of auxiliary charge
variables, which form doublets of the external $SU(2)$ symmetry.
As expected, the amplitudes computed here by exploiting the
$\widehat{\mathcal H}_6$ current algebra, coincide with the ones
resulting from the Penrose limit, {\it i.e.} the contraction, of the
amplitudes on $AdS_3\times S^3\times {\mathcal M}$. This allows us
to identify the crucial role played by the charge variables in the
fate of holography. They become coordinates on a four-dimensional
holographic screen for the pp-wave \cite{kirpio}. Global Ward
identities represent powerful constraints on the form of the
correlation functions and we would like to argue that higher
dimensional generalizations, even in the presence of R-R fluxes
where no chiral splitting is expected to take place, should follow
the same pattern. We thus believe that some of the pathologies of
the BMN limit pointed out in the literature should rather be
ascribed to an incomplete knowledge of the scaling limit in the
computation of the relevant amplitudes. Taking fully into account
the rearrangement, technincally speaking a `Saletan contraction'
\cite{sfetsos}, of the (super)conformal generators
in a $\widehat{\mathcal H}_{2+2n}$ Heisenberg algebra is
imperative in this sense.

The plan of the present paper is as follows:

In section 2 we briefly describe the Hpp-waves whose $\s$-models
are WZNW models based on the ${\bf H}_{2+2n}$ Heisenberg groups
and then we concentrate on the six-dimensional wave that
emerges from the Penrose limit of $AdS_3 \times S^3$ discussing
the corresponding
contraction of the $\widehat{SL}(2,\mathbb{R})_{k_1}\times
\widehat{SU}(2)_{k_2}$ currents. In Section 3 we identify the
relevant representations of $\widehat{\cal H}_6^L \times
\widehat{\cal H}_6^R$ and write down the explicit expressions for
the tachyon vertex operators. In section 4 we compute two and
three-point correlation functions on the world-sheet and compare
the results with those obtained from the limit of $AdS_3\times
S^3$. In section 5 we compute four-point correlation functions on
the world-sheet by means of current algebra techniques. In section
6 we present the Wakimoto free-field approach and check
consistency of the results obtained in this way with those in the
previous sections. In section 7 we study string amplitudes in the
Hpp-wave and analyze the structure of their singularities.
In section 8 we propose a concrete holographic formula relating the Hpp-wave
S-matrix elements
to precise limits of arbitrary boundary CFT correlators.
Finally we draw our conclusions and indicate lines for future
investigation.

\renewcommand{\theequation}{\arabic{section}.\arabic{equation}}
\section{Hpp-waves and the Penrose limit of $AdS_3\times S^3$}
\label{HppPenrose}

The plane wave backgrounds we will discuss in
this paper have the simple form \cite{ppwaves} \be
\label{ppntrans} ds^{2} = - 2 du dv - {1\over 4} du^{2}
\sum_{\a=1}^{n} \mu_\a^2 y_{\a}\bar{y}_{\a} + \sum_{\a=1}^{n}
dy_{\a}d\bar{y}_{\a} + \sum_{i=1}^{24-2n}g_{ij} dx^{i} dx^{j} \
\ee Here $u$ and $v$ are light-cone coordinates, $y_\a = r_\a
e^{i \f_\a}$ are complex coordinates parameterizing the $n$
transverse planes and  $x^{i}$ are the remaining $24-2n$
dimensions of the critical bosonic string that we assume
compactified on some internal manifold ${\cal M}$ with metric
$g_{ij}$. In the following we will concentrate on the $2+2n$
dimensional part of the metric in Eq. $(\ref{ppntrans})$. The wave
is supported by a non-trivial NS-NS antisymmetric tensor field
strength (whence the name Hpp-wave)
\be
H = \sum_{\a=1}^{n}\mu_\a du\wedge dy_\a\wedge d \bar y_\a \ ,
\label{atfs} \ee while the dilaton is constant and all the other
fields are set to zero.

The background defined in $(\ref{ppntrans})$ and $(\ref{atfs})$
with generic $\m_\a$ has a $(5n+2)$-dimensional isometry group
generated by translations in $u$ and $v$, independent rotations in
each of the $n$ transverse planes and $4n$ ``magnetic translations''.
When
$2 \le k \le n$ of the $\m_\a$ coincide the isometry group is enhanced:
the generic
$U(1)^n$ rotational symmetry of the metric is enlarged to $SO(2k)
\times U(1)^{n-k}$, broken to $U(k) \times
U(1)^{n-k}$ by the field strength of the antisymmetric tensor. The
dimension of the resulting isometry group is therefore
$5n+2+k(k-1)$.

As first realized in \cite{nw} for the case $n=1$ and then in
\cite{kehag} for generic $n$, the $\s$-models corresponding to
Hpp-waves are WZNW models based on the ${\bf H}_{2+2n}$ Heisenberg
group. The $\widehat{\mathcal H}_{2+2n}$ current algebra is
defined by the following OPEs \ba P^+_{\alpha}(z) {P}^{-\beta}(w)
&\sim& {2\delta_{\alpha}^{\beta} \over (z-w)^{2}}
- {2 i \m_\a \delta_{\alpha}^{\beta}\over (z-w)}
K(w)\nonumber \ , \\
J(z) P^+_{\alpha}(w) &\sim& -{i \m_\a \over (z-w)}P^+_{\alpha}(w) \ , \nb
\\
J(z) {P}^{-\alpha}(w) &\sim& +{i \m_\a \over (z-w)}
{P}^{-\alpha}(w) \ , \nb \\
J(z)K(w) &\sim& \frac{1}{(z-w)^2} \ ,
\label{opes}  \ea where $\a, \b = 1, ..., n$.
The anti-holomorphic currents satisfy a similar set of OPEs
\footnote{As usual we will distinguish the right objets by putting
a bar on them.} and the total affine symmetry of the model is
$\widehat{\mathcal H}^L_{2+2n} \times \widehat{\mathcal H}^R_{2+2n}$.

A few clarifications are in order. First of all the zero modes of
the left and right currents only realize a $(4n+3)$-dimensional
subgroup of the whole isometry group. The left and right central
elements  \footnote{Notice that we use the same letter for both
a (spin $s$) current $W(z)$ and the corresponding charge
$W\equiv W_0=\oint
\frac{dz}{2\pi i}\,z^{s-1} W(z) $. In order to avoid any
confusion  we try  always to emphasize the two-dimensional nature
of the world-sheet fields by showing their explicit $z$
dependence.} $K$ and $\bar K$ are identified and generate
translations in $v$; $P^+_\a$ and $P^{-\a}$ together with
their right counterparts generate the $4n$ magnetic translations;
$J+\bar J$ generates translations in $u$ and $J - \bar J$ a
simultaneous rotation in all the $n$ transverse planes.
In the following we will refer to the subgroup of the isometry group
that is not generated by the zero modes of the currents as $G_I$.

The position of the index $\a = 1, ..., n$ carried by the $P^\pm$
generators is meant to emphasize that at the point where the
generic $U(1)^n$ part of the isometry group is enhanced to $U(n) =
SU(n)_I \times U(1)_{J-\bar J}$ they transform respectively in the
fundamental and in the anti-fundamental representation of $SU(n)_I$.
The left and right current modes satisfy the same
commutation relations with the generators of the $SU(n)_I$ symmetry
of the background.

Let us discuss some particular cases. When $n=1$ we have the
original NW model and all the background isometries are realized
by the zero-modes of the currents. When $n=2$ there is an
additional $U(1)_I$ symmetry which extends to  $SU(2)_I$ for
$\m_1=\m_2$. In this paper we will describe in detail only the
six-dimensional Hpp-wave, because of its relation to the BMN limit
of the $AdS_3/CFT_2$ correspondence. Higher-dimensional Hpp-waves
do not display any new special feature. When we discuss the
Wakimoto representation for the ${\bf H}_6$ WZNW model, the
following change of variables
\be y^{\alpha} = e^{i \m_\a u/2}
w^{\alpha} \ , \hspace{1cm} \bar{y}_{\alpha}= e^{-i \m_\a u/2}
\bar{w}_{\alpha} \ , \ee which yields a metric with a $U(2)$
invariant form \be \label{pp2trans} ds^{2} = - 2 du dv + {i\over
4} du \sum_{\a=1}^{2} \m_\a( w^{\alpha}d\bar{w}_{\alpha} -
\bar{w}_{\alpha}dw^{\alpha}) + \sum_{i=1}^{2}
dw^{\alpha}d\bar{w}_{\alpha} \ . \ee will prove useful.

As it is well known, the background $(\ref{ppntrans})$, $(\ref{atfs})$
with $n=2$ and $\m_1=\m_2$ arises from the Penrose limit of $AdS_3\times
S^3$,
the near horizon geometry of an $F1|NS5$ bound state.
The general metric with $\m_1 \ne \m_2$ can also be obtained as a Penrose
limit
but starting with different curvatures for $AdS_3$ and $S^3$.
In global coordinates the metric can be written as
\be ds_6^{2} = R_1^{2} \ [
-(\cosh\rho)^{2} dt^{2} + d\rho^{2} + (\sinh\rho)^{2}
d\varphi_{1}^{2}] + R_2^2 \ [(\cos\theta)^{2} d\psi^{2} + d\theta^{2} +
(\sin\theta)^{2} d\varphi_{2}^{2}] \ , \ee where
$(t,\rho,\varphi_1)$ parameterize the three dimensional
anti-de Sitter space with curvature radius $R_1$
and $(\theta,\psi,\varphi_2)$ parameterize $S^3$ with curvature radius
$R_2$.
In the Penrose limit we focus on a  null geodesic of a particle
moving along the
axis of $AdS_3$ ($\rho \to 0$) and spinning around the equator of the
three sphere ($\theta\to 0$). We then change variables according to
\be t = \frac{\m_1 u}{2} + \frac{v}{\m_1 R_1^{2}} \qquad
\psi = \frac{\m_2 u}{2} - \frac{v}{\m_2 R^{2}} \qquad \rho =
{r_{1}\over R_1} \qquad \theta = {r_{2}\over R_2} \ , \ee and take
the limit sending $R_1, R_2 \to \infty$ while keeping $\m_1^2
R_1^2 = \m_2^2 R_2^2$.

{}From the world-sheet point of view, the Penrose limit of $AdS_3
\times S^3$ amounts to a contraction of the current algebra of the
underlying $\widehat{SL}(2,\mathbb{R})\times \widehat{SU}(2)$ WZNW
model. The $\widehat{SL}(2,\mathbb{R})$ current algebra at level
$k_1$ is given by \ba \label{sl2}
  K^+(z)K^-(w) &\sim&
\frac{k_1}{(z-w)^2} -\frac{2 K^3(w)}{z-w} \ , \nonumber\\
K^3(z)K^\pm(w) &\sim& \pm \frac{K^{\pm}(w)}{z-w} \ , \nonumber \\
K^3(z)K^3(w) &\sim& \displaystyle - \frac{k_1}{2(z-w)^2} \ .
\ea
Similarly the $\widehat{SU}(2)$ current algebra at level $k_2$ is
\ba J^+(z)J^-(w) &\sim&
\frac{k_2}{(z-w)^2} + \frac{2 J^3(w)}{z-w} \ , \nonumber\\
J^3(z)J^\pm(w) &\sim& \pm \frac{J^{\pm}(w)}{z-w} \ , \nonumber \\
J^3(z)J^3(w) &\sim& \frac{k_2}{2 (z-w)^2} \ . \label{su2} \ea

The contraction to the $\widehat{\mathcal H}_6$ algebra defined in
$(\ref{opes})$ is performed by first introducing the new currents
\ba P_1^\pm &=& \sqrt{\frac{2}{k_1}} K^\pm \ , \hspace{1.6cm}
P_2^\pm = \sqrt{\frac{2}{k_2}} J^\pm \ , \nb \\
J &=& - i (\m_1 K^3 + \m_2 J^3) \ , \hspace{0.6cm} K =
- i \left (\frac{K^3}{\m_1 k_1} - \frac{J^3}{\m_2 k_2}  \right ) \ ,
\label{contr}
\ea and then by taking the limit $k_1, k_2 \rightarrow \infty$
with $\m_1^2 k_1 = \m_2^2 k_2$.

In view of possible applications of our analysis to the
superstring, and in order to be able to consider flat space or a
torus with $c_{int}=20$ as a consistent choice for the internal
manifold ${\cal M}$ of the bosonic string before the Penrose limit
is taken, one should choose $k_{1}-2=k_{2}+2=k$ so that the
central charge is $c=6$.

\renewcommand{\theequation}{\arabic{section}.\arabic{subsection}.\arabic{equation}}
\section{Spectrum of the model}
\label{spectrum}

Our aim in this section is to determine
the spectrum of the string in the Hpp-wave with ${\bf H}_6$
Heisenberg symmetry. As in the ${\bf H}_4$ case,  in addition to
`standard' highest-weight representations, new modified
highest-weight (MHW) representations should be included. In the
${\bf H}_4$ case as well as in ${\bf H}_6$ with $SU(2)_I$
symmetry, such MHW representations are actually spectral flowed
representations. However, in the general ${\bf H}_6$ $\mu_1\neq
\mu_2$ case, we have the novel phenomenon that spectral flow
cannot generate the MHW representations.

The MHW representations are difficult to handle in the
current algebra formalism. Fortunately
they are  easy to analyze in the quasi-free field
representation \cite{kk,dak} where their unitarity and their
interactions are straightforward.

\subsection{${\mathcal H}_6$ representations}
\setcounter{equation}{0}

The representation theory of the extended Heisenberg algebras,
such as ${\mathcal H}_6$, is very similar to the ${\cal H}_4$ case
\cite{kk,dak}.
The ${\mathcal H}_6$
commutation relations are \be [P^+_\a,P^{-\b}] = -2 i \m_\a
\d_\a^\b K \ , \hspace{0.4cm} [J,P^{+}_\a] = - i \m_\a P^+_\a \ ,
\hspace{0.4cm} [J,P^{-\a}] = i \m_a P^{-\a} \ . \ee

As explained
in the previous paragraph this algebra generically admits an
additional $U(1)_I$ generator $I^3$ that satisfies \be
[I^3,P^+_\a] = -i  (\s^3)_\a{}^\b P^+_\b \ , \hspace{1cm}
[I^3,P^{-\a}] = i  (\s^{3,t})^\a{}_\b P^{-\b} \ . \ee When
$\m_1 = \m_2 \equiv \m$ the $U(1)_I$ symmetry is enhanced to
$SU(2)_I$ \be[I^{a}, P^+_{\alpha}] = - i
(\sigma^{a})_{\alpha}^{\phantom{\alpha}\beta}P^+_{\beta} \ , \qquad
[I^{a}, {P}^{-\alpha}] = i
(\sigma^{a,t})^{\alpha}{}_{\beta}{P}^{-\beta}  \ , \hspace{0.4cm} a = 1,2,3
\ .\ee

For ${\mathcal H}_6$ there are two Casimir operators: the
central element $K$ and the combination \be {\cal
C}=2JK+\frac{1}{2} \sum_{\a=1}^2 (P^+_{\alpha}P^{-\alpha}+P^{-\alpha}P^+_{\alpha})
\ . \ee

There are three types of unitary representations:

1) Lowest-weight representations $ V_{p,\jh}^+$, where $p > 0$. They are
constructed
starting from a state $|p,\jh\rangle$ which satisfies $P^+_{\alpha}|p,\jh
\rangle=0$,
$K|p,\jh\rangle = i p |p,\jh\rangle$ and $J|p,\jh\rangle = i \jh
|p,\jh\rangle$.
The spectrum of $J$ is given by $\{ \jh +\m_1 n_1+\m_2 n_2 \}$, $n_1,n_2 \in
\mathbb{N}$ and the value of the Casimir is $\mathcal{C}=-2p\jh+(\m_1+\m_2)p$ .

2) Highest-weight representations $ V_{p,\jh}^-$, where $p > 0$. They are
constructed
starting from a state $|p,\jh\rangle$ which satisfies $P^-_{\alpha}|p,\jh
\rangle=0$,
$K|p,\jh\rangle = - i p |p,\jh\rangle$ and $J|p,\jh\rangle = i \jh
|p,\jh\rangle$.
The spectrum of $J$ is given by $\{ \jh -\m_1 n_1-\m_2 n_2 \}$, $n_1,n_2 \in
\mathbb{N}$ and the value of the
Casimir is $\mathcal{C}= 2p\jh+(\m_1+\m_2)p$. The representation
$V^-_{p,-\jh}$ is conjugate to $V^+_{p,\jh}$.

3) Continuous representations $V _{s_1,s_2,\jh}^0$ with $p=0$. These
representations
are characterized by $K |s_1,s_2,\hj \rangle =  0$,
$J |s_1,s_2,\hj \rangle = i \jh|s_1,s_2,\hj \rangle $ and
$P^\pm_{\alpha}|s_1,s_2,\hj \rangle\neq 0$.
The spectrum of $J$ is then given by $\{ \jh +\m_1 n_1+\m_2 n_2 \}$, with
$n_1,n_2 \in \mathbb{Z}$ and $| \jh | \le \frac{\m}{2}$ where
$\m = {\rm min}(\m_1,\m_2)$. In this case we have two other Casimirs besides
$K$:
${\cal C}_1 = P^+_1 P^{-1}$ and ${\cal C}_2 = P^+_2 P^{-2}$. Their values
are
${\cal C}_\a = s^2_\a$, with $s_\a \ge 0$ and $\a=1, 2$.
The one dimensional  representation can be
considered as a particular continuous representation, where the
charges $s_\a$ and $\jh$ are zero.

The ground states of all these
representations are assumed to be invariant under the $U(1)_I$
($SU(2)_I$) symmetry. This follows from comparison with the
spectrum of the scalar Laplacian in the gravitational wave
background, described  below.

Since we are dealing with infinite dimensional representations, it
is very convenient to introduce charge variables in order to keep
track of the various components of a given representation in a
compact form. We introduce two doublets of charge variables $x_\a$
and $x^\a$, $\a = 1, 2$. The action of the ${\mathcal H}_6$ generators
and of the additional generator $I^3$ on the $V_{p,\jh}^+$
representations is given by \ba P_\a^+ &=& \sqrt{2} \m_\a p x_\a \
, \hspace{1cm}
P^{-\a} = \sqrt{2} \p^\a \ , \hspace{1cm} K = i p \ , \nb \\
J &=& i \left ( \jh + \m_\a x_\a \p^\a \right ) \ ,
\hspace{1cm} I^3 =i x_\a
(\s^{3,t})^\a{}_\b \p^\b \ . \label{difp} \ea
Similarly for the $V_{p,\jh}^-$ representations we have
\ba P_\a^+ &=&  \sqrt{2} \p_\a \ , \hspace{1cm}
P^{-\a} = \sqrt{2} \m_\a p x^\a\ , \hspace{1cm} K = - i p \ , \nb \\
J &=& i \left ( \jh - \m_\a x^\a \p_\a \right ) \ , \hspace{1cm}
I^3 = - i x^\a (\s^{3,t})^\a{}_\b \p^\b \ . \label{difm} \ea
Finally for the $V_{s_1,s_2,\jh}^0$ representations we have \be
P_\a^+ = s_\a x_\a \ , \hspace{0.4cm} P^{-\a} = s_\a x^\a \ ,
\hspace{0.4cm} J = i \left ( \jh + \m_\a x_\a \p^\a
\right ) \ ,  \hspace{0.4cm} I^3 =i x_\a
(\s^{3,t})^\a{}_\b \p^\b \ ,  \label{dif0} \ee with the
constraints $x^1 x_1 = x^2 x_2 = 1$, \ie \ $x_\a = e^{i\phi_\a}$.
Alternative representations of the generators are possible. In
particular, acting on $V_{s,\jh}^0$, it may prove convenient to
introduce charge variables $\xi_\alpha$ such that $\sum_{\alpha}
\xi_{\alpha}\xi^{\alpha}=1$. The $\xi_\alpha$ are related to the
$x_\alpha$ in (\ref{dif0}) by $\xi_\a = \frac{s_\a}{s} x_\a$ where
$s^2 = s_1^2+s_2^2$.

We can easily organize the spectrum of the D'Alembertian in the
plane wave background in representations of ${\mathcal H}^L_6
\times {\mathcal H}_6^R$.
Using radial coordinates in the two transverse planes
the covariant scalar D'Alembertian reads
\be
\nabla^2 = - 2\p_u \p_v + \sum_{\a=1}^2 \left ( \p_{r_\a}^2
+\frac{1}{r_\a^2} \p^2_{\f_\a}+\frac{1}{r_\a} \p_{r_\a}
+\frac{\m^2_\a}{4} r_\a^2 \p^2_v \right )  \ ,
\label{dalam}
\ee
and its scalar eigenfunctions may be taken to be of the form
\be f_{p^+,p^-}(u,v,r_\a,\f_\a) = e^{i p^+ v + i p^- u} g(r_\a,\f_\a) \ . \ee
For $p^+ \neq 0$, $g(r_\a,\f_\a)$ is given by the product of wave-functions
for two harmonic oscillators in two dimensions with frequencies
$\omega_\a = \left|p^+\right| \mu_\a/2$
\be
g_{l_\a,m_\a}(r_\a,\f_\a) = \left ( \frac{l_\a!}{2 \pi (l_\a+|m_\a|)!} \right )^{\frac{1}{2}}
e^{i m_\a \f_\a} e^{-\frac{\xi_\a}{2}} \xi_\a^{\frac{|m_\a|}{2}} L_{l_\a}^{|m_\a|}(\xi_\a) \ ,
\label{2dho}
\ee
with $\xi_\a = \frac{\m_\a p^+ r_\a^2}{2}$ and $l_\a \in \mathbb{N}$, $m_\a \in \mathbb{Z}$.
The resulting eigenvalue is
\be
\Lambda_{p^+\ne0} = 2p^+p^- - \sum_{\a=1}^2 \m_\a \left | p^+ \right | (2l_\a+|m_\a|+1) \ .
\label{daleig}
\ee
and by comparison with the value of the Casimir on the  ${\mathcal H}^L_6
\times {\mathcal H}_6^R$ representations we can identify
\be
p = \left | p^+ \right | \ , \hspace{0.68cm}
\jh =  p^- - \sum_{\a=1}^2 \m_\a(2l_\a+|m_\a|)  \ , \hspace{0.68cm}
m_\a = n_\a - \bar{n}_\a   \ , \hspace{0.68cm}
l_\a = {\rm Max}(n_\a,\bar{n}_\a)   \ .
\ee
For $p^+=0$ the $g(r_\a,\f_\a)$ can be taken to be Bessel functions
and they give the decomposition of a plane wave whose radial momentum
in the two transverse planes is $s_\a^2$, $\a=1,2$.

\subsection{$\widehat{\mathcal H}_6$ representations and long strings}
\setcounter{equation}{0}

The representations of the affine Heisenberg algebra
$\widehat{\mathcal H}_{6}$ that will be relevant for the study of
string theory in the six-dimensional Hpp-wave are the
highest-weight representations with a unitary base and some new
representations with a modified highest-weight condition that we
will introduce below and that in the case $\m_1=\m_2$ coincide
with the spectral flowed representations.

The OPEs in $(\ref{opes})$ correspond to the following commutation relations
for the $\widehat{\mathcal H}^L_6$ left-moving current modes
\be
[P^+_{\alpha\,n},P^{-\beta}_m] = 2n \delta_{\alpha}^{\beta}\,
\delta_{n+m}-2i \m_\a \delta_{\alpha}^{\beta}K_{n+m} \ , \hspace{0.8cm}
[J_n,K_m]=n \delta_{n+m,0} \ , \nb
\ee
\be
[J_n,P^+_{\alpha\,m}] = -i \m_\a P^+_{\alpha\,n+m}  \ , \hspace{2.5cm}
[J_n,P^{-\alpha}_m]=  i \m_\a P^{-\alpha}_{n+m} \ .
\ee

There are  three types of highest-weight representations. Affine
representations based on $V^\pm_{p,\jh}$ representations of the
horizontal algebra, with conformal dimension
\be h = \mp p \jh + \frac{\m_1 p}{2}(1-\m_1 p)+ \frac{\m_2
p}{2}(1-\m_2 p) \ , \ee and affine representations based on
$V^0_{s_1,s_2,\jh}$ representations, with conformal dimension \be
h = \frac{s_1^2}{2} + \frac{s_2^2}{2} = {s^2\over 2} \ . \ee

In the current algebra formalism we can introduce a doublet of charge
variables and regroup the infinite number of fields that appear in
a given representation of $\widehat{\mathcal H}^L_6$ in a single field
\be
\Phi_{p,\hj}^+(z;x_{\alpha}) = \sum_{n_1, n_2 = 0}^{\infty} \prod_{\a=1}^2
\frac{(x_\a \sqrt{\m_\a p})^{n_\a}}{\sqrt{n_\a!}} R_{p,\hj; n_1, n_2}^+(z) \
,
\quad p>0  \ , \ee
\be
\Phi_{p,\hj}^-(z;x^{\alpha}) = \sum_{n_1, n_2 = 0}^{\infty} \prod_{\a=1}^2
\frac{(x^\a \sqrt{\m_\a p})^{n_\a}}{\sqrt{n_\a!}} R_{p,\hj; n_1, n_2}^-(z) \
,
\quad p>0 \ , \ee
\be \Phi_{s_1,s_2,\hj}^0(z;x_\a) =
\sum_{n_1,n_2=-\infty}^{\infty} \prod_{\a=1}^2 (x_\a)^{n_\a}
R_{s_1,s_2,\jh; n_1,n_2}^0(z) \ , \quad s_1,s_2\geq 0 \ . \ee

Highest-weight representations of the current algebra lead to a
string spectrum free from negative norm states only if they
satisfy the constraint \be {\rm Max}(\m_1 p,\m_2 p)  < 1 \ .
\label{bound} \ee

When $\m_1 = \m_2 = \m$ new representations should be considered
that result from spectral flow of the original representations
\cite{mo1}. Spectral flowed representations are highest-weight
representations of an isomorphic algebra whose modes are related
to the original ones by \ba \tilde{P}^+_{\a, n} &=& P^+_{\a, n -
w} \ , \hspace{1cm} \tilde{P}^{- \a}_{n} = P^{- \a}_{n + w} \ ,
\hspace{1cm}
\tilde{J}_n = J_n \ , \nb \\
\tilde{K}_n &=& K_n - i w \d_{n,0} \ ,  \hspace{1cm} \tilde{L}_n =
L_n - i w J_n \ . \ea The long strings in this case can move
freely in the two transverse planes and correspond to the spectral
flowed type 0 representations, exactly as for the ${\bf H}_4$ NW
model \cite{dak}.

In the general case $\m_1 \ne \m_2$
a similar interpretation is not possible.
However instead of introducing new representations
as spectral flowed representations we can still define
them through a modified highest-weight condition.
Such Modified Highest Weight (MHW) representations are a more general
concept
compared to spectral flowed representations, as the analysis for
$\mu_1\not=\mu_2$
indicates.

In order to understand which kind of representations are
needed for the description of states with $p$ outside the range
$(\ref{bound})$, it is useful to resort to a free field
realization of the $\widehat{\mathcal H}_{2+2n}$ algebras, first
introduced for the original NW model in \cite{kk}. This representation
provides an interesting relation between primary vertex operators
and twist fields in orbifold models. For $\widehat{\mathcal H}_6$ we
introduce
a pair of free bosons  $u(z),v(z)$ with $\langle v(z)u(w) \rangle  =
\log{(z-w)}$ and two complex bosons $y_\a(z) = \xi_\alpha(z) + i
\eta_\alpha(z) $ and $\tilde{y}^\a(z) = \xi_\alpha(z) - i \eta_\alpha(z) $
with $\langle y_\a(z)\tilde{y}^\b(w) \rangle = -2 \d^\b_\a
\log{(z-w)}$. The currents \ba
J(z) &=& \p v(z) \ , \hspace{2cm} K(z) =  \p u(z) \ , \nb \\
P^+_\a(z) &=& i e^{-i \m_\a u(z)} \p y_\a(z) \ , \hspace{1cm}
P^{-\a}(z) = i e^{i \m_\a u(z)} \p \tilde{y}^\a(z) \ , \label{ff1}
\ea satisfy the $\widehat{\mathcal H}_6$ OPEs $(\ref{opes})$. The
ground state of a $V^{\pm}_{p,\jh}$ representation is given by the
primary field \be R^{\pm}_{p,\jh;0}(z) = e^{i [\jh u(z) \pm p
v(z)]} {\s}^{\mp}_{\m_1 p}(z) {\s}^{\mp}_{\m_2 p}(z) \ .
\label{ff2} \ee The $ {\s}^{\mp}_{\m p}(z)$ are twist fields,
characterized by the following OPEs \ba \p y(z) {\s}^-_{\m p}(w)
&\sim& (z-w)^{-\m p} \ {\tau}_{\m p}^-(w) \ , \hspace{1cm}
\p \tilde{y}(z) {\s}^-_{\m p}(w) \sim (z-w)^{-1+\m p} \ {\s}_{\m p}^{-
(1)}(w)\ , \nb \\
\p y(z) {\s}^+_{\m p}(w) &\sim& (z-w)^{-1+\m p} \ {\s}_{\m p}^{+
(1)}(w)\ , \hspace{0.4cm} \p \tilde{y}(z) {\s}^+_{\m p}(w) \sim
(z-w)^{-\m p} \ {\tau}_{\m p}^+(w) \ , \label{ff3} \ea where
${\tau}_{\m p}^\pm(z)$ and ${\s}_{\m p}^{\pm (1)}(z)$ are excited
twist fields.  The ground state of a $V_{s_1,s_2,\jh}^0$
representation is determined by the primary field \be
R^0_{s_1,s_2,\jh;0}(z) = e^{i \jh u(z)} R^0_{s_1}(z)R^0_{s_2}(z) \
, \label{ff4} \ee where \be R^0_{s_\a}(z) = \frac{1}{2\pi}
\int_0^{2 \pi} d \t_\a e^{ \frac{i s_\a }{2} \left ( y_\a(z) e^{-i
\t_\a} + \tilde{y}^\a(z) e^{i \t_\a}  \right )} \ . \label{ff5}
\ee
are essentially free vertex operators.

In analogy with $\widehat{\mathcal H}_4$ we define for
arbitrary $\m p > 0$ \ba R^{\pm}_{p,\jh;0}(z) &=& e^{i [\jh u(z)
\pm p v(z)]} {\s}^{\mp}_{\{\m_1 p\}}(z) {\s}^{\mp}_{\{\m_2 p\}}(z)
\ , \hspace{1cm}
\{ \m_1 p \} \ne 0 \ , \{ \m_2 p \} \ne 0 \ , \nb \\
R^{\pm}_{p,\jh,s_1;0}(z) &=& e^{i [\jh u(z) \pm p v(z)]}
R^0_{s_1}(z) {\s}^{\mp}_{\{\m_2 p\}}(z) \ , \hspace{1.4cm}
\{ \m_1 p \} = 0 \ , \{ \m_2 p \} \ne 0 \ , \nb \\
R^{\pm}_{p,\jh,s_2;0}(z) &=& e^{i [\jh u(z) \pm p v(z)]}
{\s}^{\mp}_{\{\m_1 p\}}(z) R^0_{s_2}(z) \ , \hspace{1.4cm}
\{ \m_1 p \} \ne 0 \ , \{ \m_2 p \} = 0 \ ,
\label{ff6}
\ea
where $[\m p]$ and $\{ \m p \}$ are
the integer and fractional part of $\m p$ respectively.
Quantization of the model in the light-cone gauge shows that the resulting
string spectrum is unitary.
{}From the current algebra point of view
the states that do not satisfy the bound $(\ref{bound})$
belong to new representations which satisfy a modified
highest-weight condition and are defined as follows.
When $K_0 |p,\jh \rangle = i \m p |p,\jh \rangle$ with  $\{ \m_\a p \} \ne
0$,
$\a =1, 2$, the affine representations we are interested in are defined by
\ba
P^+_{\a, \ n} |p,\jh\rangle &=& 0 \ , \hspace{0.4cm} n \ge - [\m_\a p] \ ,
\hspace{1cm}
P^{-\a}_{n} |p,\jh\rangle = 0 \ , \hspace{0.4cm} n \ge 1 + [\m_\a p] \ ,
\nb \\
J_{n} |p,\jh\rangle &=& 0 \ , \hspace{0.4cm} n \ge 1 \ , \hspace{1cm}
K_{n} |p,\jh\rangle = 0 \ , \hspace{0.4cm} n \ge 1  \ .
\label{ff7}
\ea
Similarly when $K_0 |p,\jh \rangle = - i \m p |p,\jh \rangle$ with  $\{
\m_\a p \} \ne 0$,
$\a =1, 2$, the affine representations we are interested in are defined by
\ba
P^+_{\a, \ n} |p,\jh\rangle &=& 0 \ , \hspace{0.4cm} n \ge  1 + [\m_\a p] \
, \hspace{1cm}
P^{-\a}_{n} |p,\jh\rangle = 0 \ , \hspace{0.4cm} n \ge - [\m_\a p] \ ,  \nb
\\
J_{n} |p,\jh\rangle &=& 0 \ , \hspace{0.4cm} n \ge 1 \ , \hspace{1cm}
K_{n} |p,\jh\rangle = 0 \ , \hspace{0.4cm} n \ge 1  \ .
\label{ff8}
\ea
Finally whenever either  $\{ \m_1 p \} = 0$ or   $\{ \m_2 p \} = 0$ we
introduce
new ground states $|p,s_1,\jh \rangle$ and  $|p,s_2,\jh \rangle$
which satisfy the same conditions as in $(\ref{ff7})$, $(\ref{ff8})$
except that
\be
P^+_{\a, \ n} |p,\jh,s_\a \rangle = 0 \ , \hspace{0.4cm} n \ge - [\m_\a p] \
, \hspace{1cm}
P^{- \a}_{n} |p,\jh,s_\a \rangle = 0 \ , \hspace{0.4cm} n \ge  [\m_\a p] \ ,
\label{ff9}
\ee
for either $\a =1$ or $\a=2$.

These states correspond to strings that do not feel any more the confining
potential in one of the two transverse planes. The presence of these states
in the spectrum can be justified along similar lines as for $AdS_3$
\cite{mo1} or the ${\bf H}_4$ \cite{kirpio,dak} WZNW models.

\renewcommand{\theequation}{\arabic{section}.\arabic{subsection}.\arabic{equation}}
\section{Three-point functions}
\label{3point}

We now turn to compute the simplest interactions in the
Hpp-wave, encoded in the three-point functions of the scalar
(tachyon) vertex operators identified in the previous section. We
will initially discuss the non symmetric $\mu_1\neq \mu_2$ case,
where global Ward identities can be used to completely fix the
form of the correlators. We will then address the $SU(2)_I$
symmetric case and argue that the requirement of non-chiral
$SU(2)_I$ invariance is crucial in getting a unique result.
We will finally describe the derivation of the two and
three-point functions starting from the corresponding
quantities in $AdS_3 \times S^3$.

\subsection{${\cal H}_6$ three-point couplings}
\setcounter{equation}{0}

In the last section we have seen that the primary fields of the $\widehat
{\mathcal H}^L_6\times \widehat
{\mathcal H}^R_6$ affine algebra are of the form
\beq
\Phi_{\n}^{a}(z,\bz;x,\bx) \ ,
\eeq
where $a={\pm,0}$ labels the type of representation and
$\n$ stands for the charges that are necessary in order to
completely specify the representation, {\it i.e}. $\n=(p,\hj)$ for
$V^{\pm}$ and $\n=(s_1,s_2,\hj)$ for $V^0$. Finally
$x$ stands for the charge variables we introduced to keep track of the
states that form a given representation: $x = x_\a$ for ${V}^{+}$,
$x = x^\a$ for ${V}^{-}$ and  $x = x_\a$ with
$x_\a = 1/x^\a$ ({\it i.e.} $x_\a = e^{i\phi_\a}$) for ${V}^{0}$ .
In the following we will leave the dependence of the vertex operators
on the anti-holomorphic variables $\bar z$ and $\bar x$ understood.
The OPE between the currents and the primary vertex operators can be
written in a compact form
\be
{\cal J}^A(z) \F^a_\n(w;x) = {\cal D}^A_a \frac{\F^a_\n(w;x)}{z-w} \ ,
\ee
where $A$ labels the six  $\widehat{\mathcal H}_6$ currents
and the $ {\cal D}^A_a$ are the differential operators that realize
the action of ${\cal J}^A_0$ on a given representation $(a,\n)$,
according to $(\ref{difp})$, $(\ref{difm})$ and $(\ref{dif0})$.

We fix the normalization of the operators in
the ${V}^{\pm}_{p_1,\hj_1}$ representations by choosing the overall
constants in their two-point functions, which are not determined by
the world-sheet or target space symmetries, to be such that \be
\langle\Phi^+_{p_1,\hj_1}(z_1,x_{1\alpha})\Phi^-_{p_2,\hj_2}(z_2,x_2^{\alpha})
\rangle = \frac{|\prod_{\a=1}^2 e^{-p_1 \m_\a
x_{1\alpha}x_2^{\alpha}}|^2}
{|z_{12}|^{4h}}\delta(p_1-p_2)\delta(\hj_1+\hj_2) \ , \label{pm} \ee where
we
introduced the shorthand notation $f(z,x)f(\bz,\bx)=|f(z,x)|^2$.
Similarly, the other non-trivial two-point functions are chosen to
be \be \langle\Phi^0_{s_{1\alpha},\hj_1}(z_1,x_{1\alpha})
\Phi^0_{s_{2\alpha},\hj_2}(z_2,x_{2\alpha})\rangle =
\prod_{\alpha=1,2}{\delta(s_{1\alpha}- s_{2\alpha}) \over
s_{1\alpha}} \delta(\phi_{1\alpha} - \phi_{2\alpha})
\delta(\bar\phi_{1\alpha} - \bar\phi_{2\alpha}) \d(\jh_1+\jh_2) \
, \ee where we set $x_{i\alpha} = e^{i\phi_{i\alpha}}$.

Three-point functions, denoted by $G_{abc}(z_i,x_i)$ or more
simply by $\langle a b c\rangle$ in the following, are determined
by conformal invariance on the world-sheet to be of the form \be
\label{three-point} \langle\Phi_{\n_1}^a(z_1,x_1)\Phi_{\n_2}^b(z_2,x_2)
\Phi_{\n_3}^c(z_3,x_3)\rangle=
\frac{C_{abc}(\n_1,\n_2,\n_3)K_{abc}(x_1,x_2,x_3)}
{|z_{12}|^{2(h_1+h_2-h_3)}|z_{13}|^{2(h_2+h_3-h_2)}|z_{23}|^{2(h_2+h_3-h_1)}}
\ , \ee where $C_{abc}$ are the quantum structure constants of the
CFT and  the `kinematical' coefficients $K_{abc}$ contain all the
dependence on the ${\mathcal H}^L_6\times {\mathcal H}^R_6$ charge
variables $x$ and $\bx$. For generic values of $\m_1$ and $\m_2$
($\frac{\m_1}{\m_2} \notin \mathbb{Q}$), the functions $K_{abc}$
are completely fixed by the global Ward identities, as it was the
case for the ${\bf H}_4$ WZNW model \cite{dak}. When $\m_1 = \m_2$
we will have to impose the additional requirement of $SU(2)_I$
invariance. An important piece of information for understanding
the structure of the three-point couplings is provided by the
decomposition of the tensor products between representations of
the ${\mathcal H}_6$ horizontal algebra
\ba V^+_{p_1,\jh_1} \otimes V^+_{p_2,\jh_2} &=&
\sum_{n_1,n_2=0}^\infty V^+_{p_1+p_2,\jh_1+\jh_2+\m_1 n_1+\m_2n_2}
\ , \nb \\
V^+_{p_1,\jh_1} \otimes V^-_{p_2,\jh_2} &=&
\sum_{n_1,n_2=0}^\infty V^+_{p_1+p_2,\jh_1+\jh_2-\m_1 n_1-\m_2n_2}
\ , \hspace{0.5cm}  p_1 > p_2  \ , \nb \\
V^+_{p_1,\jh_1} \otimes V^-_{p_2,\jh_2} &=&
\sum_{n_1,n_2=0}^\infty V^-_{p_1+p_2,\jh_1+\jh_2+\m_1 n_1+\m_2n_2}
\ , \hspace{0.5cm} p_1 < p_2 \ .  \label{tensorH6} \ea
Note that when $\m_1=\m_2$ there are $n+1$ terms with the same
$\jh = \jh_1+\jh_2 \pm \m n$ in
$(\ref{tensorH6})$. The existence of this multiplicity
is precisely what is necessary in order to obtain $SU(2)_I$
invariant couplings, as we will explain in the following.
We will also need
\ba
V^+_{p \, ,\jh_1} \otimes V^-_{p \, ,\jh_2} &=& \int_0^\infty s_1 ds_1
\int_0^\infty s_2 ds_2 V^0_{s_1,s_2,\jh_1+\jh_2} \ , \nb \\
V^+_{p_1,\jh_1} \otimes V^0_{s_1,s_2,\jh_2} &=&
\sum_{n_1,n_2=-\infty}^\infty V^+_{p_1+p_2,\jh_1+\jh_2+\m_1 n_1+\m_2n_2}
\ . \label{tensorH6b} \ea

Let us first discuss the generic case $\m_1 \ne \m_2$, starting
from $\langle++-\rangle$. According to $(\ref{tensorH6})$ this
coupling is non-vanishing only when $p_1+p_2 = p_3$ and $L =
-(\jh_1+\jh_2+\jh_3) = \m_1 q_1 + \m_2 q_2$, with $q_1,q_2\in
\mathbb{N}$. The global Ward identities can be unambiguously
solved and the result is\footnote{The standard $\delta$-function
for the Cartan conservation rules are always implied. We do not
write them explicitly. } \be K_{++-}(q_1,q_2) = \left |
\prod_{\a=1}^2 e^{-\m_\a
x^\a_3(p_1x_{1\a}+p_2x_{2\a})}(x_{2\a}-x_{1\a})^{q_\a} \right |^2
\label{cgppm}\ . \ee

The corresponding three-point couplings are \be {\it
C}_{++-}(q_1,q_2) = \prod_{\a=1}^2\frac{1}{q_\a!} \left [
\frac{\g(\m_\a p_3)}{\g(\m_\a p_1)\g(\m_\a p_2)} \right
]^{\frac{1}{2}+q_\a} \ , \label{qpppm}  \ee
where $\gamma(x) = \Gamma(x)/ \Gamma(1-x)$. All other
couplings that only involve $\F^\pm$ vertex operators follow from
$(\ref{cgppm})$, $(\ref{qpppm})$ by permutation of the indices and
by using the fact that $K_{++-}{C}_{++-} \rightarrow
K_{--+}{C}_{--+}$ up to the exchange $x_i^\a \leftrightarrow
x_{i\a}$ and the inversion of the signs of all the $\jh_i$.

Similarly the $\langle +-0 \rangle$ coupling can be non-zero only
when $p_1=p_2$ and $L = -(\jh_1+\jh_2+\jh_3) = \sum_\alpha \m_\a
q_\a$, with $q_\a \in \mathbb{Z}$. Global Ward identities yield

\be K_{+-0} = \left | \prod_{\a=1}^2 e^{- \m_\a p_1
x_{1\a}x_{2}^\a-\frac{s_\a}{\sqrt{2}} \left (x_2^\a
x_{3\a}+x_{1\a} x_3^\a \right)} x_{3\a}^{q_\a} \right |^2 \ .
\label{cpmo1}  \ee  Moreover \be
{C}_{+-0}(p,\jh_1;p,\jh_2;s_1,s_2,\jh_3) = \prod_{\a=1}^2
e^{\frac{s_\a^2}{2}[\psi(\m_\a p)+\psi(1-\m_\a p)-2\psi(1)]} \ ,
\label{cpmo2} \ee where $\psi(x) = \frac{d \ln{\G(x)}}{dx}$ is the
digamma function.

Finally the coupling between three $\F^0$ vertex operators simply
reflects momentum conservation in the two transverse planes.
Therefore it is non-zero only when  \be s_{3\a}^2 =
s_{1\a}^2+s_{2\a}^2+2s_{1\a}s_{2\a} \cos{\xi_{\a}} \ ,
\hspace{1cm} s_{3\a}e^{i\h_{\a}} = -s_{1\a}-s_{2\a}e^{i \xi_\a} \
, \hspace{0.6cm} \a = 1, 2 \ , \label{conserv} \ee where
$\xi_\a=\phi_{2\a}-\phi_{1\a}$ and $\h_\a=\phi_{3\a}-\phi_{1\a}$.
It can be written as \be K_{000}(\phi_{1\a},\phi_{2\a},\phi_{3\a})
= \prod_{\a=1}^2 \frac{ 8 \pi^2 \d(\xi_\a+\bar{\xi}_\a)\d(\h_\a +
\bar{\h}_\a)}{\sqrt{4s_{1\a}^2s_{2\a}^2-(s_{3\a}^2-s_{1\a}^2-s_{2\a}^2)^2}}
e^{-iq_\a(\phi_{1\a}+\bar \phi_{1\a})} \ , \label{cooo} \ee where the
angles $\xi_\a$ and  $\h_\a$ are fixed by the Eqs. (\ref{conserv})
and again $L= \sum_\a \m_\a q_\a$ with $q_\a \in \mathbb{Z}$.

As discussed in section \ref{HppPenrose}, when $\m_1=\m_2=\m$ the
plane wave background displays an additional $SU(2)_I$
symmetry. At the same time we see from $(\ref{tensorH6})$ that
there are also new possible couplings and they precisely combine
to give an $SU(2)_I$ invariant result. Let us start again from
three-point couplings containing only $\F^\pm$ vertex operators.
In this case the $SU(2)_I$ invariant result is obtained after
summing over all the couplings $C_{++-}(q_1,q_2)$ with
$(q_1+q_2)=L/\mu = Q$ \ba
\label{su2ppm}
K_{++-}(Q){\it C}_{++-}(Q) &=&
\sum_{q_1=0}^{Q}
K_{++-}(q_1,Q-q_1) C_{++-}(q_1,Q-q_1)  \\
&=& \frac{1}{Q!}\left [ \frac{\g(\m
p_3)}{\g(\m p_1)\g(\m p_2)} \right ]^{\frac{1}{2}+Q}\left |e^{-\m
\sum_{\a=1}^2 x_3^\a(p_1x_{1\a}+p_2x_{2\a})} \right |^2
||x_{2}-x_{1}||^{2Q} \ , \nb
\ea where $||x||^2 \equiv \sum_{\a}
|x_\a|^2$ is indeed $SU(2)_I$ invariant.

Similarly the $\langle +-0 \rangle$ correlator becomes, after
summing over $q_1 \in \mathbb{Z}$ , \ba & &
K_{+-0}(Q){C}_{+-0}(p,\jh_1;p,\jh_2;s_1,s_2,\jh_3) =
\prod_{\a=1}^2  \left | e^{- \m p_1
x_{1\a}x_{2}^\a-\frac{s_\a}{\sqrt{2}} \left (x_2^\a
x_{3\a}+x_{1\a} x_3^\a \right)} \right |^2
\left ( \frac{||x_{3}||^2}{2} \right )^{Q} \nb \\
& &
e^{\frac{s_1^2+s_2^2}{2}[\psi(\m p)+\psi(1-\m p)-2\psi(1)]} \ ,
\label{cpmo2} \ea
with the constraint $x_{31}\bar{x}_3^1 = x_{32}\bar{x}_3^2$.
The $\langle 000 \rangle$ coupling gets
similarly modified.

\subsection{The Penrose limit of the charge variables}
\setcounter{equation}{0}

It is interesting to discuss how the three-point couplings in the
Hpp-wave with ${\mathcal H}^L_6\times {\mathcal H}^R_6$ symmetry
are related to the three-point couplings in $AdS_3 \times S^3$. The first
thing we have to understand is
how the ${\mathcal H}_6$ representations arise in
the limit from representations of $SL(2,\mathbb{R}) \times SU(2)$. For
$SU(2)$ we have the representations ${V}(\tilde{l})$ with $2\tilde{l}
\in \mathbb{N}$ and $\tilde{m} = -\tilde{l}, -\tilde{l}+1, ..., \tilde{l}$.
For $SL(2,\mathbb{R})$ we have
three types of unitary normalizable representations:

1) Lowest-weight discrete representations ${\cal D}^+(l)$,
constructed starting form a state $|l \rangle$ which
satisfies $K^- |l \rangle = 0$, with $l
>1/2$. The spectrum of $K^3$ is given by $\{l+n\}$, $n \in
\mathbb{N}$ and the Casimir is ${\cal C}_{SL} = -l(l-1)$.

2) Highest-weight discrete representations ${\cal D}^-(l)$,
constructed starting form a state $|l \rangle$ which
satisfies $K^+ |l \rangle = 0$, with $l
>1/2$. The spectrum of $K^3$ is given by $\{-l-n \}$, $n \in
\mathbb{N}$ and the Casimir is ${\cal C}_{SL} = -l(l-1)$.

3) Continuous representations ${\cal D}^0(l,\a)$, constructed
starting form a state $|l,\a \rangle$ which satisfies $K^\pm |l,\a
\rangle \ne 0$, with $l = 1/2 + i \s $, $\s \ge 0$. The spectrum
of $K^3$ is given by $\{\a+n \}$, $n \in \mathbb{Z}$ and $0 \le \a
<1$. The Casimir is ${\cal C}_{SL} = 1/4 + \s^2$.

Let us start with the $V^+_{p,\jh}$ representations.
Following \cite{dak} we consider states that sit near  the top of an
$SU(2)$ representation
\be
\tilde{l} = \frac{k_2}{2} \m_2 p - b  \ , \hspace{1cm} \tilde{m}
=  \frac{k_2}{2} \m_2 p - b - n_2  \ .
\ee
In order to get in the limit states with a finite conformal dimension and
well defined quantum numbers with respect to the currents in
$(\ref{contr})$,
we have to choose for $SL(2,\mathbb{R})$ a ${\cal D}^-(l)$ representation with
\be
l = \frac{k_1}{2} \m_1 p - a \ , \hspace{1cm} m = -\frac{k_1}{2} \m_1 p + a
-n_1  \ .
\ee
In the limit $\jh = - \m_1 a + \m_2 b$.
Reasoning in a similar way one can see that the $V^-_{p,\jh}$
representations result from ${\cal D}^+(l) \times {V}(\tilde{l})$
representations with
\ba
l &=& \frac{k_1}{2} \m_1 p - a \ , \hspace{1cm}
m = \frac{k_1}{2} \m_1 p - a + n_1  \ , \nb \\
\tilde{l} &=& \frac{k_2}{2} \m_2 p - b  \ , \hspace{1cm} \tilde{m}
=  - \frac{k_2}{2} \m_2 p + b + n_2  \ , \ea and  $\jh = \m_1 a -
\m_2 b$ in the limit. Finally the $V^0_{s_1,s_2,\jh}$
representations result from ${\cal D}^0(l,\a) \times
{V}(\tilde{l})$ representations with \be l = \frac{1}{2} + i
\sqrt{\frac{k_1}{2}}s_1 \ , \hspace{0.4cm} m = \a + n_1  \ ,
\hspace{0.4cm} \tilde{l} = \sqrt{\frac{k_2}{2}}s_2  \ ,
\hspace{0.4cm} \tilde{m} = n_2 \ , \ee and $\jh = - \m_1 \a$. The
tensor product of these representations reproduces in the limit
the ones displayed before for ${\mathcal H}_6$ in Eq.
$(\ref{tensorH6})$.

Let us briefly discuss  how the Penrose limit acts on the
wave-functions corresponding to the representations considered above.
We will consider only the limit of the ground states
but the analysis can be easily extended to the limit of the whole
$SL(2,\mathbb{R}) \times SU(2)$ representation if we introduce
a generating function for the matrix elements, which can be expressed
in terms of the Jacobi functions.

Using global coordinates for $AdS_3 \times S^3$,
the ground state of a ${\cal D}^-_l \times V(\tilde{l})$ representation
can be written as
\be
e^{2 i l t - 2 i \tilde{l} \psi} ({\rm cosh}\r)^{-2l}
({\rm cos}\theta)^{2 \tilde{l}} \ .
\ee
After scaling the coordinates and the quantum numbers as required
by the Penrose limit this function becomes
\be
e^{2 i p v + i \jh u -\frac{p}{2}(\m_1r^2_1+\m_2r^2_2)} \ , \hspace{1cm}
\jh = -\m_1 a + \m_2 b \ .
\ee
In the same way starting from a ${\cal D}^+_l \times V(\tilde{l})$ representation
\be
e^{-2 i l t + 2 i \tilde{l} \psi} ({\rm cosh}\r)^{-2l}
({\rm cos}\theta)^{2 \tilde{l}} \ ,
\ee
we obtain
\be
e^{-2 i p v + i \jh u -\frac{p}{2}(\m_1r^2_1+\m_2r^2_2)} \ , \hspace{1cm}
\jh = \m_1 a - \m_2 b \ .
\ee
As anticipated, the limit of the generating functions lead
to semiclassical wave-functions for the six-dimensional wave
which are a simple generalizations of those
displayed in \cite{dak}.

We introduce a vertex operator for each unitary
representations of $SL(2,\mathbb{R})$
\ba
\Psi^+_l(z,x) &=& \sum_{n=0}^\infty c_{l,n} (-x)^{n} R^+_{l,n}(z) \ , \nb \\
\Psi^-_l(z,x) &=& \sum_{n=0}^\infty c_{l,n} x^{-2l-n} R^-_{l,n}(z) \ , \nb \\
\Psi^0_{l,\a}(z,x) &=& \sum_{n \in \mathbb{Z}} x^{-l+\a+n} R^0_{l,\a,n}(z)
\ , \ea where $c_{l,n}^2 = \frac{\G(2l+n)}{\G(n+1)\G(2l)}$. The
differential operators that represent the $SL(2,\mathbb{R})$ action are
\be {\cal D}_1^- = - x^2 \p_x - 2 l x \ , \hspace{0.4cm}  {\cal D}_1^+ = - \p_x \ ,
\hspace{0.4cm}  {\cal D}_1^3 = l + x \p_x \ . \ee
Similarly
for $S^3$ we introduce \be \Omega_{\tilde{l}}(z,y) =
\sum_{m=-\tilde{l}}^{\tilde{l}}
\tilde{c}_{\tilde{l},m} y^{\tilde{l}+m} R_{\tilde{l},m}(z) \ ,  \\
\ee where $\tilde{c}_{\tilde{l},m}^2 =
\frac{\G(2\tilde{l}+1)}{\G(\tilde{l}+m+1)\G(\tilde{l}-m+1)}$ and
the differential operators that represent the $SU(2)$ action
are \be {\cal D}_2^+ =
\p_y \ , \hspace{0.4cm} {\cal D}_2^- = -y^2 \p_y + 2 \tl y \ ,
\hspace{0.4cm} {\cal D}_2^3 = y \p_y - \tl \ . \ee Generalizing the case
studied in \cite{dak}, we can now implement the Penrose limit on
the operators $\Psi^a_l(z,x) \Omega_{\tilde{l}}(z,y)$ and
determine their precise relation with the $\hat {\cal H}_6$
operators $\Phi^a(z,x,y)$. In this section we shall denote the two
${\cal H}_6$ charge variables as $x$ and $y$ in order to emphasize
that they are related to the charge variables of
$SL(2,\mathbb{R})$ and $SU(2)$ respectively.
For the discrete representations we have
\ba \label{pp+} \Phi^+_{p,\hj}(z,x,y) &=&
\lim_{k_1, k_2 \to\infty} \left(\frac{x}{\sqrt{k_1}}\right)^{-2l}
\left(\frac{y}{\sqrt{k_2}}\right)^{2\tl}
\Psi^-_{l}\left(z,\frac{\sqrt{k_1}}{x}\right) \ \Omega_{\tilde l}
\left(z,\frac{\sqrt{k_2}}{ y}\right) \ ,  \\
\label{pp-} \Phi^-_{p,\hj}(z,x,y) &=& \lim_{k_1,k_2\to\infty}
\Psi^+_l\left(z,-\frac{x}{\sqrt{k_1}}\right) \Omega_{\tl}
\left(z,\frac{y}{\sqrt{k_2}}\right) \ , \ea with \be l =
\frac{k_1}{2} \m_1 p - a \ , \hspace{1cm} \tilde{l} =
\frac{k_2}{2} \m_2 p - b \ . \ee
For the continuous representations we have
\be \label{pp0}
\Phi^0_{s_1,s_2,\hj}(z,x,y) = \lim_{k_1,k_2\to\infty} (-i
x)^{-l+\a} \, y^{\tilde{l}} \ \Psi^0_{l,\a}\left(z, \frac{i}{x} \right)
\Omega_{\tl} \left(z,\frac{1}{y}\right) \ , \ee with \be
l = \frac{1}{2} + i\sqrt{\frac{k_1}{2}} s_1 \ , \hspace{1cm}
\tilde{l} = \sqrt{\frac{k_2}{2}} s_2 \ . \ee

With the help of the previous formulae it is not difficult to find
the Clebsch-Gordan coefficients of the plane-wave three-point
correlators. In fact, a similar analysis has been performed in
\cite{dak} for the three-point correlators of the Nappi-Witten
gravitational wave considered as a limit of $SU(2)_k\times U(1)$.
For $AdS_3$ the
general form of the three point function is fixed, up to
normalization, by $\widehat {SL}(2,\mathbb{R})_L \times
\widehat{SL}(2,\mathbb{R})_R$  invariance ($x$ dependence) and by
$SL(2,\mathbb{C})$ global conformal invariance on the world-sheet ($z$
dependence), to be \beq \label{CGsl2}
\left \langle\,\prod_{i=1}^3\Psi_{l_1}(z_i,\bz_i,x_i,\bx_i)\, \right \rangle
=C(l_1,l_2,l_3)\prod_{i<j}^{1,3}
\frac{1}{|x_{ij}|^{2l_{ij}}|z_{ij}|^{2h_{ij}}}\, , \eeq where
$l_{12}=l_1+l_2-l_3$, $h_{12}=h_1+h_2-h_3$ and cyclic permutation
of the indexes. Due to the $\widehat{SU}(2)_L \times
\widehat{SU}(2)_R$ and world-sheet conformal invariance the
correlation function of three primaries on  $S^3$ is given by \beq
\label{CGsu2}
\left \langle\,\prod_{i=1}^3\Omega_{\tl_i}(z_i,\bz_i,y_i,\by_i)\, \right \rangle
=C(\tl_1,\tl_2,\tl_3)\prod_{i<j}^{1,3}\frac{|y_{ij}|^{2\tl_{ij}}}{|z_{ij}|^{2h_{ij}}}\,
, \eeq where $\tl_{ij}$ and $h_{ij}$ are defined as for
(\ref{CGsl2}).
Let us consider for instance the limit leading to a
$\langle ++-\rangle$ correlator. Taking into account that $\sum_i
\hj_i = -L = - \mu_1 (a_1+a_2-a_3) + \mu_2 (b_1 + b_2 - b_3)$, the
kinematic coefficient receives the following contribution from the
$AdS_3$ part \beq K_{++-}(x,\bx)= k_1^{-q_1} \left
|e^{-\mu_1 x_{3}(p_1x_1+p_2x_2)}\right|^2 |x_2-x_1|^{2 q_1} \ , \eeq
where $q_1 = a_1+a_2-a_3$ and a similar contribution from the
$S^3$ part \beq K_{++-}(y,\by)= k_2^{-q_2}\left
|e^{-\mu_2 y_{3}(p_1y_1+p_2y_2)}\right|^2 |y_2-y_1|^{2 q_2} \ , \eeq
where $q_2 = - b_1-b_2+b_3$. Putting the two contributions
together \beq K_{++-}(x,\bx,y,\by)=  k_1^{-q_1}
k_2^{-q_2} \left |e^{-\mu_1
x_{3}(p_1x_1+p_2x_2)}e^{-\mu_2 y_{3}(p_1y_1+p_2y_2)}\right|^2
|x_2-x_1|^{2 q_1}|y_2-y_1|^{2 q_2} \ , \eeq we reproduce
$(\ref{cgppm})$. In the $SU(2)$ invariant case $\mu_1=\mu_2$, one
finds a looser constraint on the $a_i$ and $b_i$ that leads to
$q_1+q_2= Q = -L/\mu$. Summing over the allowed values of $q_1$
and $q_2$ one eventually gets the $SU(2)_I$ invariant result
$(\ref{su2ppm})$.
Using the above expression for the CG coefficients for
a coupling of the form $\langle +-0 \rangle $ one
obtains \beq K_{+-0} (x,\bx,y,\by) =
\left |e^{-\mu_1p_1x_1 x_2- \frac{s_{1}}{\sqrt{2}}(x_2 x_3+x_1x_{3})} \right |^2
\left |e^{-\mu_2 p_1 y_1 y_2- \frac{s_{2}}{\sqrt{2}}(y_2 y_3+y_1y_{3})} \right |^2
|x_3|^{2q_1} |y_3|^{2q_2} \ , \eeq where $q_1 = a_1-a_2+\a$ and
$q_2 = b_2-b_1$.

\subsection{The Penrose limit of the  $AdS_3 \times S^3$ three-point couplings}
\setcounter{equation}{0}

We now turn to the Penrose limit of the $AdS_3 \times S^3$ structure constants.
The limit of the $SU(2)$ three-point couplings \cite{zf} has been
considered in \cite{dak} and we refer to that paper for a detailed discussion.
Here we provide a similar analysis for the $SL(2,\mathbb{R})$
structure constants \cite{tesch}
and show that when combined with the $SU(2)$
part they reproduce in the limit the $\hat{\cal H}_6$ structure constants.

In general, the $AdS_3/CFT_2$ correspondence entails the exact
equivalence between (super)string theory on $AdS_3\times {\cal
K}$, where ${\cal K}$ is some compact space represented by a unitary
CFT on the worldsheet, and a CFT defined on the boundary of
$AdS_3$. Equivalence at the quantum level implies a
isomorphism of the Hilbert spaces and of the operator algebras
of the two theories.
For various reasons it is often convenient to consider the
Euclidean version of $AdS_3$ described by an $SL(2,\mathbb{C})/SU(2)$ WZNW
model on the hyperbolic space $H^+_3$ with $S^2$ boundary.
Although the Lorentzian $SL(2,\mathbb{R})$ WZNW model and the Euclidean
$SL(2,\mathbb{C})/SU(2)$ WZNW model are formally related by analytic
continuation of the string coordinates, their spectra are quite
distinct. As observed in \cite{mo1, mo3}, except for unflowed
($w=0$) continuous representations, physical string states on
Lorentzian $AdS_3$ corresponds to non-normalizable states in the
Euclidean $SL(2,\mathbb{C})/SU(2)$ model. Yet unitarity of the dual boundary
$CFT_2$ that follows from positivity of the Hamiltonian and slow
growth of the density of states should make the analytic continuation
legitimate. Indeed correlation functions for the Lorentzian
$SL(2,\mathbb{R})$ WZNW model have been obtained by analytic continuation
of those for the Euclidean $SL(2,\mathbb{C})/SU(2)$ WZNW model
\cite{mo3}. Singularities displayed by correlators involving
non-normalizable states have been given a physical interpretation
both at the level of the worldsheet, as due to worldsheet
instantons, and of the target space. Some singularities have been
associated to operator mixing and other to the non-compactness of
the target space of the boundary $CFT_2$. The failure of the
factorization of some four-point string amplitudes has been given
an explanation in \cite{mo3} and argued not to prevent the
validity of the analytic continuation from Euclidean to Lorentzian
signature. Since we are going to take a Penrose limit of $SL(2,\mathbb{R})$
correlation functions computed by analytic continuation from
$SL(2,\mathbb{C})/SU(2)$, we need to assume the validity of this procedure.
Reversing the argument, the agreement we found between correlation
functions in the Hpp-wave computed by current algebra techniques
or by the Wakimoto representation with those resulting from the
Penrose limit (current contraction) of the $SL(2,\mathbb{C})/SU(2)$ WZNW
model should be taken as further evidence for the validity of the
analytic continuation.

For the euclidean $AdS_3$, that is the $H_3^+$ WZNW model,
the two and three-point functions involving vertex operators in
 unitary representations were computed by Teschner \cite{tesch}.
The two-point functions are given by
\be
\langle \Psi_{l_1}(x_1,z_1) \Psi_{l_2}(x_2,z_2) \rangle
= \frac{1}{|z_{12}|^{4h_{l_1}}} \left [
\frac{\d^2(x_1-x_2) \d(l_1+l_2-1)}{B(l_1)}
+\frac{\d(l_1-l_2)}{|x_{12}|^{4 l_1}} \right ] \ ,
\ee
where
\be
B(l) = \frac{\n^{1-2l}}{\pi b^2 \g(b^2(2l-1))}  \ ,
\hspace{0.4cm} \n = \pi \frac{\G(1-b^2)}{\G(1+b^2)} \ ,
\hspace{0.4cm} b^2 = \frac{1}{k_1-2} \ ,
\ee
and $l = \frac{1}{2} + i \s$.
The three-point functions have the same dependence on the $z_i$ and the
$x_i$ as displayed in $(\ref{CGsl2})$. The structure constants are
given by
\ba
\label{sl2qcg}
C(l_1,l_2,l_3) &=&  - \frac{b^2 Y_b(b)}{2 \sqrt{\pi \n} \g(1+b^2)}
\prod_{i=1}^3 \frac{\sqrt{\g(b^2(2l_i-1))}}{G_b(1-2l_i)} \times \\
&\times&
G_b(1-l_1-l_2-l_3)G_b(l_3-l_1-l_2)G_b(l_2-l_1-l_3)G_b(l_1-l_2-l_3)  \ . \nb
\ea
In the previous expression we used the entire function $Y_b(z)$
introduced in \cite{zz} and the closely related function
$G_b(z)$ given by
\be G_b(z) = \frac{b^{-b^2z \left ( z+1+\frac{1}{b^2} \right
)}}{Y_b(-bz)} \ . \label{gtoy} \ee
The function $Y_b$ satisfies \be Y_b(z+b) = \g(bz)Y_b(z)b^{1-2bz} \ ,
\hspace{1cm} Y_b(z) = Y_b(b+1/b-z) \ . \ee
In order to study the Penrose limit of the $SL(2,\mathbb{R})$
structure constants we express the function $G_b(z)$ in term
of the function $P_b(z)$
that appears in the $SU(2)$ three-point functions \cite{zf} and whose
asymptotic behaviour was studied in \cite{dak}.
For this purpose we write
\be \ln
P_b(z) = f(b^2,b^2|z)-f(1-zb^2,b^2|z) \ , \ee where $f(a,b \, |z)$ is
the Dorn-Otto function \cite{do} and then use the relation
\be
f(bu,b^2|z) - f(bv,b^2|z) = \ln Y_b(v) - \ln Y_b(u) +zb(u-v) \ln b \ ,
\hspace{1cm} u+v = b + \frac{1}{b} - zb \ .
\ee
The result is
\be G_b(z) =
\frac{b \g(-b^2 z)}{Y_b(b)P_b(-z)} \ ,
\ee
and we can rewrite the  coupling $(\ref{sl2qcg})$ using the function $P_b$
\ba
\label{sl2qcp}
C(l_1,l_2,l_3) &=& - \frac{b^3}{2 \sqrt{\pi \n} \g(1+b^2)}
\prod_{i=1}^3 \frac{P_b(2l_i-1)}{\sqrt{\g(b^2(2l_i-1))}} \times \\
&\times&
\frac{\g(b^2(l_1+l_2+l_3-1))\g(b^2(l_1+l_2-l_3))\g(b^2(l_1+l_3-l_2))\g(b^2(l_2+l_3-l_1))}
{P_b(l_1+l_2+l_3-1)P_b(l_1+l_2-l_3)P_b(l_1+l_3-l_2)P_b(l_2+l_3-l_1)}  \ . \nb
\ea
Let us consider first the $\langle ++- \rangle$ coupling. As
we explained before, the $AdS_3$ quantum numbers have to be scaled as follows
\be l_i =
\frac{k_1}{2} \m_1 p_i - a_i \ .
\ee
The leading behaviour is
\be
C(l_1,l_2,l_3) \sim \ \frac{1}{2 \pi b q_1} \frac{1}{P_b(-q_1)}
\left [ \frac{\g(\m_1 p_3)}{\g(\m_1 p_1)\g(\m_1 p_2)} \right ]^{\frac{1}{2}+q_1} \ ,
\ee
where $q_1 = a_1+a_2-a_3$.
Due to the presence of $P_b(-q_1)$ in the denominator, the coupling
vanishes unless $q_1 \in \mathbb{N}$, thus reproducing the
classical tensor products (\ref{tensorH6}).
We can then write
\be
\lim_{b \to 0} C(l_1,l_2,l_3) = (-1)^{q_1}
\frac{k_1^{q_1+\frac{1}{2}}}{q_1!}
\left [ \frac{\g(\m_1 p_3)}{\g(\m_1 p_1)\g(\m_1 p_2)} \right ]^{\frac{1}{2}+q_1}
\sum_{n \in \mathbb{N}} \d(q_1-n) \ .
\ee
The sign $(-1)^{q_1}$ does not appear in the ${\cal H}_6$ couplings,
a discrepancy which
might be due to some difference between the charge variables used in
\cite{tesch} and the charge variables used in the present paper.
The same limit for the $SU(2)$ three-point couplings leads to
\be
\lim_{\tilde{b} \to 0} \tilde{C}(\tl_1,\tl_2,\tl_3) =
\frac{k_2^{q_2+\frac{1}{2}}}{q_2!}
\left [ \frac{\g(\m p_3)}{\g(\m p_1)\g(\m p_2)} \right ]^{\frac{1}{2}+q_2}
\sum_{n \in \mathbb{N}} \d(q_2-n) \
\ee
where $\tilde{b}^{-2} = k_2+2$ and $q_2 = -b_1-b_2+b_3$. We then
reproduce the coupling in $\ref{qpppm}$.
Proceeding in a similar way for a $\langle + - 0 \rangle$ correlator
we obtain from $AdS_3$
\be
\lim_{b \to 0} C(l_1,l_2,l_3) = \frac{2^{- i s_1 \sqrt{2k_1}}}{\sqrt{2 \pi}}
e^{\frac{s_1^2}{2}(\psi(p) + \psi(1-p) - 2 \psi(1) )} \ ,
\ee
and similarly  from $S^3$
\be
\lim_{\tilde{b} \to 0}  \tilde{C}(\tl_1,\tl_2,\tl_3)
= \frac{2^{1+s_2 \sqrt{2k_2}}}{\sqrt{2 \pi}}
e^{\frac{s_2^2}{2}(\psi(p) + \psi(1-p) - 2 \psi(1) )} \ .
\ee

\renewcommand{\theequation}{\arabic{section}.\arabic{subsection}.\arabic{equation}}
\section{Four-point functions}
\label{4point}

Four-point correlation functions of worldsheet primary operators
are computed in this section by solving the relevant Knizhnik -
Zamolodchikov (KZ) equations. As we will explain the resulting amplitudes
are a simple generalization of the amplitudes of the ${\bf H}_4$ WZNW model.
In section \ref{wakimoto} the same
results will be reproduced by resorting to the Wakimoto free-field
representation. As in the previous section we find it convenient
to first discuss the non-symmetric ($\mu_1\neq\mu_2$) case and
then pass to the symmetric ($\mu_1=\mu_2$) case where $SU(2)_I$
invariance is needed in order to completely fix the correlators.

In general, world-sheet conformal invariance and global Ward
identities allow us to write \be G(z_i, \bar{z}_i, x_{i},
\bar{x}_{i}) = \prod_{i<j}^4 |z_{ij}|^{2 \left
(\frac{h}{3}-h_i-h_j \right) } K(x_{i},\bar{x}_{i}){\cal
G}(z,\bar{z}, x, \bar{x}) \ , \label{f1} \ee where
$h=\sum_{i=1}^4h_i$ and the $SL(2,\mathbb{C})$ invariant cross-ratios $z$,
$\bar{z}$ are defined according to \be z =
\frac{z_{12}z_{34}}{z_{13}z_{24}} \ , \hspace{1cm} \bar z =
\frac{\bar{z}_{12}\bar{z}_{34}} {\bar{z}_{13}\bar{z}_{24}} \ .
\label{f2} \ee The form of the function $K$ and the expression of
the $\widehat{\mathcal H}_6$ invariants $x$ in terms of the
$x_i$ are fixed by the global symmetries but are different for
different types of correlators and therefore their explicit form
will be given in the next sub-sections.

The four-point
amplitudes are non trivial only when \be L = - \sum_{i=1}^4 \jh_i
= \m_1q_1 + \m_2 q_2 \ , \label{f3} \ee for some integers $q_\a$.
In the generic case for a given $L$ these integers are uniquely
fixed and the Ward identities fix the form of the functions $K$ up to
a function of two ${\mathcal H}_6$ invariants\footnote{Sometimes
we will collectively denote the ${\mathcal H}_6$ invariants $x_1$
and $x_2$ by $x_\a$ with $\a=1,2$. They should not be confused
with the components of the charge variables $x_{i\a}$ that carry
an additional label associated to the insertion point $i=1, ..., 4$.}
$x_1$ and $x_2$. The KZ equations can be schematically written in
the following form \be \p_z {\cal G}(z,x_1,x_2) = \sum_{\a=1}^2
D_{{\mathcal H}_4,q_\a}(z,x_\a){\cal G}(z,x_1,x_2) \ , \label{f4}
\ee where the $D_{{\mathcal H}_4,q_\a}$ are differential operators
closely related to those that appear in the KZ equations for the NW
model based on the $\widehat{\mathcal H}_4$ affine algebra
\cite{dak}. The equations are therefore easily solved by setting
\be {\cal G}_{q_1,q_2}(z,x_1,x_2) = {\cal G}_{{\mathcal
H}_4,q_1}(z,x_1){\cal G}_{{\mathcal H}_4,q_2}(z,x_2) \ .
\label{f5} \ee

When $\m_1=\m_2$, there are several integers that satisfy
$(\ref{f3})$ and the $SU(2)_I$ invariant correlators can be
obtained by summing over all possible pairs $(q_1,q_2)$ such that
$(q_1+q_2)=L/\mu = Q$ \be {\cal G}_{Q}(z,x_1,x_2) = \sum_{q_1=0}^Q
{\cal G}_{{\mathcal H}_4,q_1}(z,x_1){\cal G}_{{\mathcal
H}_4,Q-q_1}(z,x_2) \ . \label{f6} \ee
This is the same procedure we used for the three-point functions and
reflects the existence of new couplings between states
in $\hat{\cal H}_6$ representations at the
enhanced symmetry point.  In the following we will
describe the various types of four-point correlation functions.

\subsection{$\langle +++- \rangle$ correlators}
\setcounter{equation}{0}

Consider a correlator of the form \be G_{+++-} = \langle
\F^+_{p_1,\jh_1} \F^+_{p_2,\jh_2} \F^+_{p_3,\jh_3}
\F^-_{p_4,\jh_4} \rangle \ , \hspace{1cm} p_1+p_2+p_3 = p_4 \ .
\label{ex1} \ee This is the simplest `extremal' case. {}From the
decomposition of the tensor products of ${\mathcal H}_6$
representations displayed in Eq. (\ref{tensorH6}) it follows that
the correlator vanishes for $L < 0$ while for $L \ge 0$, $L = \m_1
q_1 + \m_2 q_2$  it decomposes into the sum of a finite number $N=
(q_1+1)(q_2+1)$ of conformal blocks which correspond to the propagation
in the $s$-channel of the representations
$\F^+_{p_1+p_2,\jh_1+\jh_2+\m_1 n_1+\m_2 n_2}$ with $n_1 = 0, ...,
q_1$ and $n_2 = 0,..., q_2$. Global ${\mathcal H}_6$ symmetry
yields \be K(q_1,q_2) =\prod_{\a=1}^2\left|e^{-\m_\a
x_{4}^\a(p_1x_{1\a}+p_2x_{2\a}+p_3x_{3\a})} \right |^2
|x_{3\a}-x_{1\a}|^{2q_\a}  \label{ex2} \ , \ee up to a function of
the two invariants ($\alpha =1,2$) \be x_\a =
\frac{x_{2\a}-x_{1\a}}{x_{3\a}-x_{1\a}} \ . \ee We decompose
the amplitude in a sum over the conformal blocks and write
\be {\cal G}_{q_1,q_2}(z, \bar
z, x_\a,\bar{x}^\a) \sim  \sum_{n_1=0}^{q_1} \sum_{n_2=0}^{q_2}
{\cal F}_{n_1,n_2}(z,x_\a) \bar{{\cal F}}_{n_1,n_2}
(\bar{z},\bar{x}^\a) \ . \label{ex3} \ee We set ${\cal
F}_{n_1,n_2} = z^{\ka_{12}}(1-z)^{\ka_{14}}F_{n_1,n_2}$ where \ba
\ka_{12} &=& h_1+h_2-\frac{h}{3}-\jh_2p_1-\jh_1p_2-(\m_1^2+\m_2^2)p_1p_2 \ ,
  \\
\ka_{14} &=&
h_1+h_4-\frac{h}{3}-\jh_4p_1+\jh_1p_4+(\m_1^2+\m_2^2)p_1p_4
-(\m_1+\m_2)p_1+L(p_2+p_3) \ , \nb \label{ex4} \ea
and where the $F_{n_1,n_2}$ satisfy the following KZ equation

\ba && \p_z F_{n_1,n_2}(z,x_1,x_2) = \frac{1}{z}
\sum_{\a=1}^2 \m_\a \left [ -(p_1x_\a+p_2x_\a(1-x_\a))\p_{x_\a}
+q_\a p_2x_\a \right ]
F_{n_1,n_2}(z,x_1,x_2) \nb \\
&-& \frac{1}{1-z} \sum_{\a=1}^2 \m_\a \left [
(1-x_\a)(p_2x_\a+p_3)\p_{x_\a}+q_\a p_2(1-x_\a) \right ]
F_{n_1,n_2}(z,x_1,x_2) \ . \label{ex5} \ea

The explicit form of the conformal blocks is
\be F_{n_1,n_2}(z,x_1,x_2) = \prod_{\a=1}^2
f(\m_\a,z,x_\a)^{n_\a} g(\m_\a,z,x_\a)^{q_\a-n_\a} \ ,
\hspace{0.6cm} n_\a = 0, ..., q_\a  \ . \label{ex6} \ee
Here \ba
f(\m_\a,z,x_\a) &=& \frac{\m_\a
p_3}{1-\m_\a(p_1+p_2)}z^{1-\m_\a(p_1+p_2)}\f_0(\m_\a) - x_\a
z^{-\m_\a(p_1+p_2)}\f_1(\m_\a) \ , \nb \\
g(\m_\a,z,x_\a) &=& \g_0(\m_\a) -\frac{x_\a p_2}{p_1+p_2}
\g_1(\m_\a) \ , \label{ex7} \ea and \ba \f_0(\m) &=& F(1-\m
p_1,1+\m p_3,2-\m p_1-\m p_2,z) \ , \hspace{0.6cm} \f_1(\m) =
F(1-\m p_1,\m p_3,1-\m p_1-\m p_2,z) \ ,  \nb \\ \g_0(\m) &=& F(\m
p_2,\m p_4,\m p_1+\m p_2,z) \ , \hspace{0.6cm} \g_1(\m) = F(1+\m
p_2,\m p_4,1+\m p_1+\m p_2,z) \ , \label{ex8} \ea where
$F(a,b,c,z)$ is the standard $_1F_2$ hypergeometric function.

We can now reconstruct the four-point function as a monodromy
invariant combination of the conformal blocks and the result is
\be {\cal G}_{q_1,q_2}(z,\bar{z},x_\a, \bar{x}^\a) = |z|^{2
\ka_{12}}|1-z|^{2 \ka_{14}} \prod_{\a=1}^2
\frac{\sqrt{\tau(\m_\a)}}{q_\a!}
\left [ C_{12}(\m_\a)|f(\m_\a,z,x_\a)|^2+C_{34}(\m_\a)|g(\m_\a,z,x_\a)|^2
\right ]^{q_\a}
\ ,
\label{cor-pppm}
\ee where $\tau(\m) =C_{12}(\m)C_{34}(\m) $ and \be
C_{12}(\m) = \frac{\g(\m(p_1+p_2))}{\g(\m p_1)\g(\m p_2)} \ ,
\hspace{1cm} C_{34}(\m) = \frac{\g(\m p_4)}{\g(\m p_3)\g(\m
(p_4-p_3))} \ .
\ee
When $\m_1=\m_2=\mu$ we set $Q=L/\mu=\sum_{\a}q_{\a}$ and find the
$SU(2)_I$ invariant combination

\ba && {\it K}_Q(x_\a,\bar{x}^\a)
{\cal G}_Q(z,\bar{z},x_\a,\bar{x}^\a) = \sum_{q_1=0}^{Q}{\it
K}(q_1,Q-q_1){\cal G}_{q_1,Q-q_1}(z,\bar{z},x_\a, \bar{x}^\a)
=\nb\\
=&&|z|^{2 \ka_{12}}|1-z|^{2 \ka_{14}} \times
\times \prod_{\a=1}^2\left|e^{-\m_\a
x_{4}^\a(p_1x_{1\a}+p_2x_{2\a}+p_3x_{3\a})} \right |^2
\frac{\tau(\m)}{Q!}\times\\
&\times&
\left [ \sum_{\a=1}^2 \left (
C_{12}(\m)|x_{13\a}f(\m,z,x_\a)|^2+C_{34}(\m)|x_{13\a}g(\m,z,x_\a)|^2
\right ) \right ]^Q \ . \nb \ea

\subsection{$\langle +-+- \rangle$ correlators}
\setcounter{equation}{0}

The next class of correlators we want to discuss is of the following form
\be G_{+-+-} = \langle
\F^+_{p_1,\jh_1} \F^-_{p_2,\jh_2} \F^+_{p_3,\jh_3}
\F^-_{p_4,\jh_4} \rangle \ , \hspace{1cm} p_1+p_3=p_2+p_4 \ ,
\label{pm1} \ee and also in this case we write $L=-\sum_i
\jh_i=\sum_\alpha \m_\alpha q_\alpha $. The Ward identities give
\be K(q_1,q_2) = \prod_{\a=1}^2 \left | e^{-\m_\a p_2
x_{1\a}x^\a_2-\m_\a p_3x_{3\a}x^\a_4 -\m_\a(p_1-p_2)x_{1\a}x^\a_4}
(x_{1\a}-x_{3\a})^{q_\a} \right |^2 \ , \label{pm3} \ee and the
two invariants (no sum over $\alpha =1,2$) \be x_\a =
(x_{1\a}-x_{3\a})(x^\a_2-x^\a_4) \ . \label{pm4} \ee
We pass to the conformal blocks and set  ${\cal F}_{n_1,n_2} =
z^{\ka_{12}}(1-z)^{\ka_{14}}F_{n_1,n_2}$ where
\ba
\ka_{12} &=& h_1+h_2-\frac{h}{3}+(\m_1^2+\m_2^2)p_1p_2-\jh_2p_1+\jh_1p_2
-(\m_1+\m_2)p_2 \ ,  \nb \\
\ka_{14} &=&
h_1+h_4-\frac{h}{3}+(\m_1^2+\m_2^2)p_1p_4-\jh_4p_1+\jh_1p_4-(\m_1+\m_2)p_4
\ . \label{pm5} \ea
The $F_{n_1,n_2}$ solve the following KZ equation
\ba
&& z(1-z) \p_z F_{n_1,n_2}(z,x_1,x_2) =  z \sum_{\a=1}^2
\left [ -2a_\a x_\a \p_{x_\a} +\frac{x_\a}{4}(b_\a^2-c_\a^2)
-\right.\nb\\
&-&\left.
\r_{\a 12} - \r_{\a 14} \right ]   F_{n_1,n_2}(z,x_1,x_2)+ \sum_{\a=1}^2
\left [ x_\a\p^2_{x_\a} +\left ( a_\a x_\a+1+q_\a \right )\p_{x_\a}\right. \\
&+&\left.\frac{x_\a}{4}(a_\a^2-b_\a^2) + \r_{\a 12} \right ] F_{n_1,n_2}(z,x_1,x_2)
\ ,\nb
\label{kz-pmpm-nf}
\ea
with
\be
\r_{\a 12} = \frac{(1+q_\a)}{2}(a_\a-b_\a) \ ,  \hspace{1cm}
\r_{\a 14} = \frac{(1+q_\a)}{2}(a_\a-c_\a)\ ,
\label{pm99}
\ee
and
\be
2a_\a = \m_\a (p_1+p_3) \ , \hspace{0.5cm} b_\a = \m_\a (p_1-p_2)
\ , \hspace{0.5cm} c_\a= \m_\a (p_2-p_3) \ . \label{pm6} \ee
The conformal blocks are very similar to the conformal
blocks for the ${\bf H}_4$ WZNW model \cite{dak}
\be
F_{n_1,n_2}(z,x_1,x_2) = \prod_{\a=1}^2
\n_{n_\a} \frac{e^{\m_\a x_\a z p_3 -z(1-z) \m_\a \p \ln{f_1(\m_\a,z)}}}
{(f_1(\m_\a, z))^{1+q_\a}}
L^{q_\a}_{n_\a}[x_\a  g(\m_\a,z)]
\left ( \frac{f_2(\m_\a,z)}{f_1(\m_\a,z)} \right )^{n_\a} \ ,
\label{pm98}
\ee
where $n_\a \in \mathbb{N}$ and $L^{q}_{n}$ is the n-th generalized Laguerre
polynomial.
We also introduced the functions
\ba
f_1(\m,z) &=& F(\m p_3,1-\m p_1,1-\m p_1+\m p_2,z) \ , \nb \\
f_2(\m,z) &=& z^{\m (p_1-p_2)}F(\m p_4,1-\m p_2,1-\m p_2+\m p_1,z)
\ , \label{pm8} \ea
and
\be
g = -z(1-z) \p \ln{\left ( f_2/f_1 \right )} \ , \hspace{1cm}
\n_{n_\a} = \frac{n_\a!}{[\m_\a(p_1-p_2)]^{n_\a}} \ .
\ee
The four-point correlator can be written in a
compact form using the combination
\be S(\m_\a,z,\bar{z}) =
|f_1(\m_\a,z)|^2 - \rho(\m_\a) |f_2(\m_\a,z)|^2 \ , \hspace{1cm}
\rho(\m) = \frac{ \tilde{C}_{12}(\m) \tilde{C}_{34}(\m)}{\m^2(p_1-p_2)^2} \ ,
\label{pm2}
\ee
where we defined
\be  \tilde{C}_{12}(\m) = \frac{\g(\m p_1)}{\g(\m
p_2)\g(\m(p_1-p_2))} \ , \hspace{1cm} \tilde{C}_{34}(\m) = \frac{\g(\m
p_4)}{\g(\m p_3)\g(\m(p_4-p_3))} \ . \label{pm9} \ee
The four-point function reads
$$ {\cal
G}_{q_1,q_2}(z,\bar{z},x_\a, \bar{x}^\a) = |z|^{2\ka_{12}}
|1-z|^{2\ka_{14}}\prod_{\a=1}^2
\frac{\tau(\m_\a,q_\a)}{S(\m_\a,z)} \left |e^{ \m_\a p_3 x_\a
z-x_\a z(1-z)\p_z \ln{S(\m_\a,z)}} \right |^2\times
$$
\be \times
|x_\a z^{b_\a}(1-z)^{c_\a}|^{-q_\a}I_{q_\a}(\zeta_\a) \ ,
\label{pm12} \ee where $I_{q}(\zeta)$ is a modified
Bessel function and
\be \zeta_\a = \frac{2\sqrt{\rho(\m_a)}|\m_\a (p_1-p_2) x_\a
z^{b_\a}(1-z)^{c_\a}|}{S(\m_\a, z)} \ , \hspace{1cm} \tau(\m,q) =
 \tilde{C}_{12}(\m)^{\frac{1-q}{2}}  \tilde{C}_{34}(\m)^{\frac{1+q}{2}} \ .
\label{pm11} \ee
When $\m_1=\m_2=\mu$ the $SU(2)_I$ invariant correlator is given
by the sum over $q_1 \in \mathbb{Z}$ with $q_2 = Q - q_1$ and
$Q=L/\mu$. The addition formula for Bessel functions leads to \be
{\cal G}_Q(z,\bar{z},x_\a, \bar{x}^\a) = \frac{\tau(\m,Q)
|z|^{2\ka_{12}- b Q }|1-z|^{2\ka_{14}- c Q }}{S(\mu,z)^2}
\frac{||x_{13}||^{Q}}{||x_{24}||^{Q}} \left |e^{x z [\mu p_3 -
(1-z) \p_z \ln{S}(\mu,z)]} \right |^2 I_{Q}(\zeta) \ , \label{pmsu2inv}
\ee where \be \zeta =
\frac{2\sqrt{C_{12}C_{34}}|z^b(1-z)^c|}{S(\mu,z)} ||x_{13}||
||x_{24}|| \ , \ee and $x= x_{13}\cdot x_{24} = \sum_{\a}
(x_{1\a}-x_{3\a})(x^\a_2-x^\a_4)$ as well as $||x_{ij}||^2 =
\sum_{\a} |x_{i\a} - x_{j\a}|^2$ are $SU(2)_I$ invariant.

The factorization properties of these correlators can be analyzed following
\cite{dak}. In this way one can check that the modified highest weight
representations introduced in section \ref{spectrum} actually
appear in the intermediate channels.

\subsection{$ \langle ++- \ 0 \rangle$ correlators}
\setcounter{equation}{0}

Let us describe now a correlator of the form \be G_{++-0} = \langle
\F^+_{p_1,\jh_1} \F^+_{p_2,\jh_2}
\F^-_{p_3,\jh_3}\F^0_{s_1,s_2,\jh_4} \rangle \ , \hspace{1cm}
p_1+p_2 = p_3 \ . \label{po1} \ee {}From the global symmetry
constraints we derive \be K(q_1,q_2) = \prod_{\a=1}^2 \left |
e^{-\mu_\a x^\a_3(p_1x_{1\a}+p_2x_{2\a})- \frac{s_\a}{\sqrt{2}}
x^\a_3 x_{4\a} - \frac{s_\a}{2\sqrt{2}}(x_{1\a}+x_{2\a})x^\a_4}
x_{4\a}^{q_\a}\right |^2  \ , \label{po2} \ee
up to a function of the two invariants (no sum over
$\alpha=1,2$) \be x_\a = (x_{1\a}-x_{2\a})x^\a_4 \ .\ee
We rewrite the
conformal blocks as \be {\cal F}_{n_1,n_2} =
z^{\ka_{12}}(1-z)^{\ka_{14}} F_{n_1,n_2} \ , \ee where \ba
\ka_{12} &=& h_1+h_2-\frac{h}{3} -p_1\jh_2-p_2\jh_1-(\m_1^2+\m_2^2)p_1p_2 \
, \nb \\
\ka_{14} &=& h_1+h_4-\frac{h}{3}  -p_1\jh_4 - L p_1 -
\frac{s_1^2+s_2^2}{4} \ . \label{po4} \ea
The KZ equation then reads
\ba
&& z(1-z) \p_z F_{n_1,n_2}(z,x_1,x_2) = -\sum_{\a=1}^2
\left[ \m_\a p_3 x_\a\p_{x_\a} + \frac{s_\a}{2\sqrt{2}}\m_\a (p_1-p_2)x_\a
\right ]
F_{n_1,n_2}(z,x_1,x_2) \nb \\
&+& z \sum_{\a=1}^2 \left [ \left ( \m_\a p_2 x_\a - \frac{s_\a}{\sqrt{2}}
\right )
\p_{x_\a} - \frac{s_\a \m_\a p_2}{2\sqrt{2}}x_\a \right ]
F_{n_1,n_2}(z,x_1,x_2) \ ,
\label{kz-ppmo}
\ea
and the solution is
\be
F_{n_1,n_2}(z,x_\a) = \prod_{\a=1}^2[s_\a \f(\m_\a,z)+x_\a
\omega(\m_\a,z)]^{n_\a}
e^{s_\a^2 \h(\m_\a, z) +s_\a x_\a \chi(\m_\a,z)}
\ , \label{po5} \ee with $n_1, n_2 \ge 0$. We have introduced
the following functions
\ba
\f(\m,z) &=& \frac{z^{1-\m p_3}}{\sqrt{2}(1-\m p_3)}F(1-\m p_1,1-\m p_3,2-\m
p_3,z) \ , \nb \\
\omega(\m,z) &=& -z^{-\m p_3}(1-z)^{\m p_1} \ , \nb \\
\chi(\m,z) &=& -\frac{1}{2\sqrt{2}} +\frac{p_2}{\sqrt{2}p_3}(1-z)F(1+\m
p_2,1,1+\m p_3,z) \ , \nb \\
\h(\m,z) &=& -\frac{z p_2}{2p_3} \ {}_3F_2(1+\m p_2,1,1;1+\m p_3,2;z)  -
\frac{1}{4} \ln{(1-z)} \ . \label{po6} \ea
The four-point function is then given by \ba {\cal
G}_{q_1,q_2}(z,\bar{z},x_\a, \bar{x}^\a) &=&
|z|^{2\ka_{12}}|1-z|^{2\ka_{14}} \prod_{\a=1}^2
C^{1/2}_{12}(\m_\a) C_{+-0}(\m_\a,p_3,s_\a) \times
\nb \\
& & e^{C_{12}(\m_\a)|s_\a \f(\m_\a,z)+x_\a\omega(\m_\a,z)|^2}
\left |e^{s_\a^2 \h(\m_\a,z) + s_\a x_\a\chi(\m_\a,z)} \right |^2 \ ,
\label{po7} \ea
where
\be C_{12}(\m) = \frac{\g(\m(p_1+p_2))}{\g(\m p_1)\g(\m p_2)} \ ,
\hspace{0.6cm}
C_{+-0}(\m,p_3,s) = e^{\frac{s^2}{2}[\psi(\m p_3)+\psi(1-\m p_3)-2\psi(1)]}
   \ .
\label{po8} \ee
The $SU(2)_I$ invariant correlator at the point $\m_1=\m_2=\mu$ is obtained
after summing
over $q_1 \in \mathbb{Z}$ with $q_1 + q_2 = Q = L/\mu$.

\subsection{$\langle +- \ 0 \ 0 \rangle$ correlators}
\setcounter{equation}{0}

The last correlator we have to consider is of the form \be \langle
\F^+_{p,\jh_1} \F^-_{p,\jh_2}\F^0_{s_{3\a},\jh_3}
\F^0_{s_{4\a},\jh_4} \rangle \ , \quad p_1 = p_2 . \label{pmo1} \ee The Ward
identities give \be K(q_1,q_2) = \prod_{\a=1}^2 \left | e^{-\m_\a
p x_{1\a}x^\a_2 - \frac{x_{1\a}}{\sqrt{2}} \left ( s_{3 \a} x_3^\a
+ s_{4 \a} x_4^\a \right )- \frac{x^\a_2}{\sqrt{2}} ( s_{3 \a}
x_{3 \a} +s_{4 \a} x_{4 \a})} x_{3 \a}^{q_\a} \right |^2   \ ,
\label{pmo2} \ee up to a function of the two invariants (no sum over
$\alpha=1,2$)
$x_\a = x^\a_3 x_{4\a}$.

We decompose this correlator around $z=1$ setting $u = 1-z$, since
the conformal blocks turn out to be simpler and rewrite them
as
\be {\cal F}_{n_1,n_2} =
z^{\ka_{12}}(1-z)^{\ka_{14}} F_{n_1,n_2} \ ,
\ee
where
\be \ka_{14} = h_1+h_4-\frac{h}{3} -p \jh_4 -
\sum_{\a=1}^2\frac{s_{4\a}^2}{2} \ ,
\hspace{1cm} \ka_{12} = \sum_{\a=1}^2 \frac{s_{3\a}^2+s_{4\a}^2}{2} -
\frac{h}{3} \ .
\label{pmo4} \ee
The KZ equation
\ba
\p_u F_{n_1,n_2}(z,x_1,x_2) &=& - \frac{1}{u} \sum_{\a=1}^2
\left [ \m_\a p x_\a \p_{x_\a}  + \frac{s_{3\a}s_{4\a} x_\a}{2} \right ]
F_{n_1,n_2}(z,x_1,x_2) \nb \\
&-& \frac{1}{1-u} \sum_{\a=1}^2\frac{s_{3\a}s_{4\a}}{2}
\left ( x_\a + \frac{1}{x_\a} \right ) F_{n_1,n_2}(z,x_1,x_2) \ ,
\label{pmo99}
\ea
has the solutions
\be F_{n_1,n_2}(u,x_\a) =
\prod_{\a=1}^2(x_\a u^{-\m_\a p})^{n_\a} e^{x_\a \omega(\m_\a, u)
+ x^\a \chi(\m_\a,u)} \ , \label{pmo5} \ee with $n_1, n_2 \in
\mathbb{Z}$, $x^\a = x_{3 \a} x_{4}^{\a}=1/x_\a$ and \be
\omega(\m,u) = - \frac{s_3s_4}{2 \m p} \ F(\m p,1,1+\m p,u) \ ,
\hspace{0.4cm} \chi(\m,u) =  - \frac{s_3s_4}{2(1-\m p)} \ u \
F(1-\m p,1,2 - \m p, u) \ . \label{pmo6} \ee

The four-point function is then given by \be {\cal
G}(u,\bar{u},x_\a,\bar{x}^\a) =  |u|^{2\ka_{12}}|1-u|^{2\ka_{14}}
\prod_{\a=1}^2 \tau(\m_\a) \left | e^{x_\a \omega(\m_\a,u) + x^\a
\chi(\m_\a, u)} \right |^2 \sum_{n_\a \in \mathbb{Z}} \left | x_\a
u^{-\m_\a p} \right |^{2n_\a} \ , \label{pmo7} \ee where $\tau(\m)
=  {C}_{+-0}(\m,p,s_{3}) {C}_{+-0}(\m,p,s_{4})$

The $SU(2)_I$ invariant correlator at the point $\m_1=\m_2=\mu$ is
obtained after summing over $q_1 \in \mathbb{Z}$ with $q_1 + q_2 =
Q = L/\mu$.

\renewcommand{\theequation}{\arabic{section}.\arabic{subsection}.\arabic{equation}}
\section{Wakimoto representation}
\label{wakimoto}

In this section we construct a free
field representation for the $\widehat {\mathcal H}_6$ algebra
starting from the standard Wakimoto realization for
$\widehat{SL}(2,\mathbb{R})$ and $\widehat{SU}(2)$ \cite{wak} and
contracting the currents of both CFTs as indicated in section 2.
Then we use this approach to compute two, three and four-point correlators
that only involve $\F^\pm$ vertex operators and reproduce the
results obtained in the previous sections.
This free field representation was introduced
by Cheung, Freidel and Savvidy \cite{cfs} and used to evaluate
correlation functions for $\widehat {\mathcal H}_4$.

\subsection{$\widehat {\mathcal H}_6$ free field representation}
\setcounter{equation}{0}

The Wakimoto representation of the
$\widehat{SL}(2,\mathbb{R})$ current algebra requires  a
pair of commuting ghost fields
$\beta_1(z)$ and $\gamma^1(z)$ (the index 1 here is a label) with propagator
$\langle\beta_1(z)\gamma^1(w)\rangle=1/(z-w)$, and a free boson $\phi(z)$
with
  $\langle\phi(z)\phi(w)\rangle=-\log (z-w) $. The
$\widehat{SL}(2,\mathbb{R})$  currents can then be written as
{\setlength\arraycolsep{2pt} \beqa \label{sl2w} K^+(z) &=&
\displaystyle-\beta_1 \  , \nonumber\\
K^-(z)&=&-\beta_1\gamma^1\gamma^1+\alpha_{+}\gamma^1\pd
\phi-k_1\pd\gamma^1 \  , \\
K^3(z) &=& \displaystyle -\beta_1\gamma^1+\frac{\alpha_{+}}{2}\pd
\phi  \  , \nonumber \eeqa} where $\alpha_+^2\equiv 2(k_1-2)$.
Similarly for $\widehat{SU}(2)$ we introduce a second pair of ghost
fields $\beta_2(z)$ and $\gamma^2(z)$ (here the index 2 is a
label) with world-sheet propagator
$\langle\beta_2(z)\gamma^2(w)\rangle=1/(z-w)$, and a free boson
$\varphi(z)$ with $\langle\varphi(z)\varphi(w)\rangle=-\log
(z-w)$. The currents are then given by
{\setlength\arraycolsep{2pt}\beqa \label{su2w} J^+(z) &=&
\displaystyle-\beta_2 \  , \nonumber\\
J^-(z)&=&\beta_2\gamma^2\gamma^2-i \alpha_{-}\gamma^2\pd
\varphi-k_2\pd\gamma^2 \  , \\
J^3(z) &=&-\beta_2\gamma^2+\frac{i \alpha_{-}}{2}  \pd \varphi \ ,
\nonumber \eeqa} where  $\alpha_-^2\equiv 2(k_2+2)$. In order to
obtain a Wakimoto realization for the $\widehat {\mathcal H}_6$
algebra, we rescale the two ghost systems \beq \label{scalw}
\beta_{\alpha}\to\sqrt{\frac{k_\a}{2}} \, \beta_{\alpha} \ ,
\hspace{1cm}
\gamma^{\alpha}\to\sqrt{\frac{2}{k_\a}}\,\gamma^{\alpha} \ , \eeq
and introduce the light-cone fields $u$ and $v$ \be \phi = -i
\sqrt{\frac{k_1}{2}} \m_1 u - \frac{i}{\sqrt{2 k_1}}
\frac{v}{\m_1} \ , \hspace{1cm} \varphi = \sqrt{\frac{k_2}{2}}
\m_2 u - \frac{1}{\sqrt{2 k_2}} \frac{v}{\m_2} \ , \ee with
$u(z)v(w)\sim \, \ln (z-w)$. We then perform the current
contraction as prescribed in $(\ref{contr})$, with the result \ba
\label{h6w}
P^+_{\alpha}(z)&=&-\beta_{\alpha} \ , \nonumber\\
P^{-\alpha}(z)&=&-2 \pd\gamma^{\alpha} -2 i \m_\a \pd
u\,\gamma^{\alpha} \ , \\
J(z)&=&i\, \m_\a \beta_{\alpha}\gamma^{\alpha}-\pd v
+\frac{\m_1^2+\m_2^2}{2}\p u \ ,
\nonumber\\
K(z)&=&-\pd u  . \nonumber
\ea
The $\hat {\cal H}_6$ stress-energy tensor follows from the limit
of $T_{SL(2,\mathbb{R})}(z)+T_{SU(2)}(z)$ and is given by
\be T(z) = \sum_{\a=1}^2
:\beta_{\alpha}(z)\pd\gamma^{\alpha}(z):+:\pd u(z)\,\pd v(z): - \frac{i}{2}
(\m_1+\m_2) \p^2 u\ ,
\ee
where the last term appears when expressing the normal ordered product of
the currents
in terms of the Wakimoto fields.

The $\F^+_{p,\jh}$ primary vertex operators similarly follow
from the $SL(2,\mathbb{R}) \times SU(2)$ primary vertex operators
in the ${\cal D}^-_l \times V_{\tilde{l}}$ representation
\be V_{l,m;\tilde l, \tilde m} = (-\g^1)^{-l-m}(-\g^2)^{\tilde{l}-\tilde{m}}
e^{\frac{2\tilde{l}}{\a_+} \phi+\frac{2i\tilde{l}}{\a_-} \varphi} \ ,
\ee
where $m$ is the eigenvalue of $K^3$.
Introducing the charge variables we can collect all the components in a
single
field
\be
\F^+_{l,\tilde{l}}(z,x_\a) = \left ( 1 + x_1 \g^1 \right )^{-2l}
\left ( 1-x_2 \g^2 \right )^{2\tilde{l}}
e^{\frac{2\tilde{l}}{\a_+} \phi+\frac{2\tilde{l}}{\a_-} i \varphi} \ ,
\ee
that in the large $k_1$, $k_2$  limit becomes, after rescaling $x_\a
\rightarrow
\frac{x_\a}{\sqrt{k_\a}}$
\be
\F^+_{p,\jh}(z,x_\a) = N(p,\jh) e^{-\sqrt{2} \m_\a p x_\a \g^\a - i p v - i
\left( \jh +
\frac{\m_1^2+\m_2^2}{2} p \right ) u }  \ . \label{wakp}
\ee
It is easy to verify that this field satisfies the correct OPEs with the
$\hat{\cal{ H}}_6$
currents and that its conformal dimension is
$h(p,\jh) = - p \jh +\frac{\m_1 p}{2}(1-\m_1 p)+\frac{\m_2 p}{2}(1-\m_2 p)$.
If we choose the normalization factor
$N(p,\jh) = (\g(\m_1 p)\g(\m_2 p))^{-1/2}$ the vertex operators
$(\ref{wakp})$
precisely reproduce the results obtained in the previous sections.

The  $\F^-_{p,\jh}$ vertex operators can be represented
using an integral transform \cite{cfs}
\beq \label{p<0w}
\Phi_{p,\hj}^-(z,x^{\alpha})= \prod_{\a=1}^2  \int\,
d^2x_{\alpha}\, \g(\m_a p)\frac{\m_\a^2 p^2}{2 \pi^2} \
e^{-\m_\a p x_{\alpha}x^{\alpha}} \Phi_{-p,\jh + \m_1 +
\m_2}^+(z,x_{\alpha}) \ .
\eeq

The Wakimoto representation can also be derived from the
$\s$-model action written in the following form
\beq S= \int \frac{d^2 z}{2 \pi} \left\{ -\pd u \bar{\pd}v +
\sum_{\a=1}^2 \left [ \beta_{\alpha}\bar{\pd}\gamma^{\alpha}
+\bar{\beta}^{\alpha}\pd\bar{\gamma}_{\alpha}-
\beta_{\alpha}\bar{\beta}^{\alpha} e^{-i \m_\a u}\right ] \right\} \,
\label{action} \ , \eeq as we will review in appendix \ref{app A}.
The non-chiral $SU(2)_I$ currents are
\beq \mathcal
J^a(z,\bz)=i\,\gamma^{\alpha}(\sigma^a)_{\alpha}^{\phantom{\alpha}\beta}\,\beta_{\beta}
\ ,
\hspace{1cm}
\qquad \bar{\mathcal J}^a(z,\bz)=-i\,\bar
\beta^{\alpha}(\sigma^a)_{\alpha}^{\phantom{\alpha}\beta}\,
\bar\gamma_{\beta} \ .
  \eeq
Using the equations of motion  \beq
\beta_{\alpha}=e^{i\m_\a u}\pd \bar{\gamma}_{\alpha} \ , \hspace{2cm}
\bar{\pd}\beta_{\alpha}=0 \ , \eeq
one can verify that they
satisfy  $\bar\pd\mathcal J^a+\pd\bar{\mathcal J}^a=0$.
Moreover their OPEs with the Wakimoto free fields are
\ba
  \mathcal
J^a(z,\bz)\gamma^{\alpha}(z,\bz) &\sim&
i\,\frac{\gamma^{\beta}(\sigma^a)_{\beta}^{\phantom{\beta}\alpha}}{z-w} \ ,
\hspace{1cm} \bar{\mathcal J}^a(z,\bz)\bar \gamma_{\alpha}(z,\bz) \sim
-i\, \frac{(\sigma^a)_{\alpha}^{\phantom{\alpha}\beta}\bar
\gamma_{\beta}}{\bz-\bar w} \ ,
\nonumber \\
\mathcal J^a(z,\bz)\beta_{\alpha}(z) &\sim&
-i\,\frac{(\sigma^a)_{\alpha}^{\phantom{\alpha}\beta}
\beta_{\beta}}{z-w} \ , \hspace{1cm} \bar{\mathcal J}^a(z,\bz)\bar
\beta^{\alpha}(\bz) \sim i\,\frac{\bar
\beta^{\beta}(\sigma^a)_{\beta}^{\phantom{\beta}\alpha}}{\bz-\bar w} \ .
\ea

\subsection{Correlators}
\setcounter{equation}{0}

In order to evaluate the correlation functions in this free-field
approach, we first integrate over the zero modes of the
Wakimoto fields using the invariant measure \be \int \,
du_0 dv_0 \prod_{\a=1}^2 d\gamma^{\alpha}_0\,d\bar{\gamma}_0^{\alpha}e^{i
\m_\a u_0}
\ . \ee The presence of the interaction term \be S_I= \sum_{\a=1}^2 S_{I\a}
= - \sum_{\a=1}^2 \int \frac{d^2w}{2 \pi}
~\beta_\a(w)\bar{\beta}^\a(\bar{w}) e^{-i \m_\a u(w,\bar{w})} \ ,
\ee in the action $(\ref{action})$ leads to the insertion in the
free field correlators of the screening operators \be
\sum_{q_1,q_2=0}^\infty \prod_{\a=1}^2 \frac{1}{q_\a!}\,\left (
\int \frac{d^2w_\a}{2\pi} \b_\a \bar{\b}^\a e^{-i \m_\a u} \right
)^{q_\a} \ . \ee

Negative powers of the screening operator are needed in
order to get sensible results for $n$-point correlation functions
other than the `extremal' ones, that only involve one
$\F_{p_n,\jh_n}^-$ vertex operator and $n-1$ $\F_{p_i,\jh_i}^+$
vertex operators. This means that the sum over $q_\a$ should
effectively runs
over all integers, $q_\a \in \mathbb{Z}$, not only the positive
ones.
An `extremal' $n$-point function can be written as \ba & &
\sum_{q_1,q_2=0}^\infty \prod_{\a=1}^2 \frac{1}{m_\a !}
\left\langle
\prod_{i=1}^{n-1}{\Phi}^+_{p_i,\hj_i}(z_i,\bz_i,x_{i\alpha},\bx^{\alpha}_i)
\Phi^+_{-p_4,\hj_4+\m_1+\m_2}(z_n,\bz_n,x_{n \alpha},\bx^{\alpha}_n)\,S_{I
\a}^{q_\a} \,
\right\rangle \ \label{genwak} \\
&=&  \delta\left(\sum_i^{n-1} p_i-p_n\right)
\prod_{i < j \ne 4} |z_i -z_j|^{-2p_i(\hj_j+\h \, p_j)-2p_j(\jh_i+\h \,
p_i)}
\prod_{i \ne n} |z_i-z_n|^{2p_n(\hj_i+\h \, p_i) - 2p_i(\jh_n-\h \,
p_n+\m_1+\m_2)}
\nb \\
&&  \sum^\infty_{q_1,q_2=0}
\delta\left(\, L
-\m_1 q_1 -\m_2 q_2 \,\right)~\prod_{\a=1}^2 R(\m_\a)
\left | \,e^{-\m_\a x_n^{\alpha}\sum_{i=1}^{n-1}  p_i x_{i\alpha}}\, \right
|^2
\frac{1}{q_\a!}\, \left ( -2 \m_\a^2 \mathcal{I}_{\a, n} \right )^{q_\a}  \ ,
\nonumber
\ea
where $L = - \sum_{i=1}^n \jh_i$,
$\h = \frac{\m^2_1+\m_2^2}{2}$  and
\be
{\cal I}_{\a, n} = \int \frac{d^2 w}{2\pi} \prod_{i=1}^{n-1}|z_i-w|^{-2
\m_\a p_i}
|z_n-w|^{2\m_\a p_n} \left | \sum_{i=1}^{n-1} \frac{p_i
x_{i\a}}{w-z_i}-
\frac{p_n x_{n \a}}{w-z_n} \right |^2 \ ,
\label{intwak}
\ee
with the constraint $p_n x_{n\a} = \sum_{i=1}^{n-1} p_i x_{i\a}$.
Finally the constant $R(\m)$, related to the
normalization of the operators in $(\ref{wakp})$, is given by
\be
R^2(\m) = \frac{\g(\m p_n)}{\prod_{i=1}^{n-1} \g(\m p_i)} \ .
\ee
In  $(\ref{genwak})$ the two $\d$-functions arise from the integration over
$u_0$ and $v_0$.
Similarly the integration over the $\g_{0\a}$ leads to four other
$\d$-functions
that constrain the integration over the $x_{n\a}$ variables and give the
exponential term.
The other terms in $(\ref{genwak})$ follow from the contraction of the free
Wakimoto fields.
Note that due to the second $\d$-function in $(\ref{genwak})$ the correlator
is non vanishing only when
$L = \m_1 q_1 + \m_2 q_2$ where $q_\a \in \mathbb{N}$.
Therefore the same structure we found before using current algebra
techniques
appears: for the generic background $\m_1 \ne \m_2$ only one term from the
double sum
in $(\ref{genwak})$ contributes while for the $SU(2)_I$ invariant wave
we have to add several terms.
Let us consider some examples. We will need the following integral
\cite{dotfat}
\ba
& &\int  d^2 t
|t-z|^{2(c-b-1)}|t|^{2(b-1)}|t-1|^{-2a} =
\frac{\pi \g(b)\g(c-b)}{\g(c)}|z|^{2(c-1)}|F(a,b,c;z)|^2  \nb \\
&-& \frac{\pi \g(c)\g(1+a-c)}{(1-c)^2\g(a)}|F(1+a-c,1+b-c,2-c;z)|^2 \ .
\label{dfintegral}
\ea
It follows from the general expression $(\ref{genwak})$
that the two-point function $\langle +- \rangle$ coincides with
$(\ref{pm})$, since only the $q_\a=0$ terms are non-vanishing.
For the $\langle ++- \rangle$ three-point coupling the integral
$(\ref{intwak})$ gives
\be
-2 \, \m_\a^2 \, {\cal I}_{\a,3} = |z_{12}|^{-2\m_\a p_3}|z_{23}|^{2\m_\a
p_1}
|z_{13}|^{2\m_\a p_2} \frac{\g(\m_\a p_3)}{\g(\m_\a p_1) \g(\m_\a p_2)}
\left | x_{1\a}-x_{2\a} \right |^2 \ ,
\ee
and the result precisely agrees with $(\ref{cgppm})$, $(\ref{qpppm})$.
When $\m_1=\m_2$ the sum over $q_\a$ reconstructs the $SU(2)_I$
invariant coupling $(\ref{su2ppm})$.

The four-point function $\langle+++-\rangle$ can be evaluated in a similar
way.
In this case
\ba
-2 \, \m_\a^2 \, {\cal I}_{\a, 4} &=&  |z_{12}|^{-2\m_\a
p_4}|z_{14}|^{-2\m_\a (p_1-p_4)}
|z_{34}|^{-2\m_\a (p_3-p_4)}|z_{24}|^{2\m_\a (p_2-p_4)} \nb \\
& & \left [ C_{12}(\m_\a) \left | x_{31\a} f(\m_\a,x_\a,z) \right |^2
+ C_{34}(\m_\a) \left | x_{31\a} g(\m_\a,x_\a,z) \right |^2 \right ] \ ,
\label{wpppm}
\ea
where the functions $f$ and $g$ are as defined in $(\ref{ex7})$ and
\be
C_{12}(\m) = \frac{\g(\m(p_1+p_2))}{\g(\m p_1)\g(\m p_2)} \ ,
\hspace{1cm} C_{34}(\m) = \frac{\g(\m p_4)}{\g(\m p_3)\g(\m
(p_4-p_3))} \ .
\ee
We find again complete agreement with $(\ref{cor-pppm})$.

Finally the correlator  $\langle+-+-\rangle$ can be obtained from  the
$\langle+++-\rangle$
correlator performing
the integral transform $(\ref{p<0w})$ of the vertex operator inserted in
$z_2$ \cite{cfs},
that is we send
$(p_2,\jh_2) \rightarrow (-p_2,\jh_2+\m_1+\m_2)$ and evaluate the $x_{2\a}$
integral.
We first rewrite
\ba
{\cal T} &\equiv&
\int d^2 x_{2 \a} \frac{\left | e^{-\m_\a p_2 x^\a_{24}} \right
|^2}{\G(q_\a+1)}
\left [ C_{12}(\m_\a) \left | x_{31\a} f(\m_\a,x_\a,z) \right |^2
+ C_{34}(\m_\a) \left | x_{31\a} g(\m_\a,x_\a,z) \right |^2 \right ]^{q_\a}
\nb \\
&=& \int d^2 x_{2 \a} \frac{\left | e^{-\m_\a p_2 x^\a_{24}} \right
|^2}{\G(q_\a+1)}
\left [ A x_{2\a}\bar{x}_{2\a} + B \bar{x}_{2\a}+\bar{B}x_{2\a}+E \right
]^{q_\a} \ ,
\ea
and then evaluate the integral using
\be
\int d^2 u  \left | e^{-u} u^t \right |^2 =
\pi (-1)^{-1-t} \g(1+t) \ ,
\ee
which is a limit of $(\ref{dfintegral})$.
The result is
\be
{\cal T} = \left | e^{\m_\a p_2  x^\a_{24} \frac{B_\a}{A}} \right |^2
\frac{|x^\a_{24}|^{-q_\a}}{2A} \left ( \frac{B\bar{B}-EA}{\m_\a^2 p_2^2}
\right )^{\frac{q_\a}{2}}
I_{q_\a}\left ( 2\m_\a p_2  \left | x^\a_{24} \right |
\sqrt{\frac{B\bar{B}-EA}{A^2}} \right )
\ ,
\label{wakt}
\ee
where $I_{q_\a}$ is a modified Bessel function of integer order and
\ba
&& \frac{\m_\a p_2  x^\a_{24} B}{A} = -\m_\a p_2 x_{1\a} x^\a_{24}
+\m_\a p_3 x_{13\a}  x^\a_{24}z -\m_\a p_2
x_{13\a}  x^\a_{24}z(1-z) \p_z \ln S(\m_\a,z,\bar{z}) \ , \nb \\
&& A = -\frac{\m^2_\a p^2_2}{\tilde{C}_{12}} |z|^{-2 \m_\a(p_1-p_2)}
S(\m_\a,z,\bar{z})\ , \hspace{1cm}
2 \m_\a p_2  \left |x^\a_{24}\right |\sqrt{\frac{B\bar{B}-EA}{A^2}} = \z_\a
\ .
\ea
The functions and constants that appear on the left-hand side of
the previous equations were
defined in $(\ref{pm8}-\ref{pm9})$ and $(\ref{pm11})$.
Combining $(\ref{wakt})$ with the rest of the
$\langle +++-  \rangle$ correlator
we obtain the $\langle +-+-  \rangle$ correlator and
also in this case the result coincides with $(\ref{pm12})$
when $\m_1 \ne \m_2$ and with $(\ref{pmsu2inv})$ when $\m_1=\m_2$.

\renewcommand{\theequation}{\arabic{section}.\arabic{equation}}
\section{String amplitudes}
\label{amplitudes}

In this section we study the string amplitudes in the Hpp-wave.
After combining the results of the previous sections with
the ones for the internal CFT and for the world-sheet ghosts, one
can easily extract irreducible vertices and decay rates in closed
form. The world-sheet integrals needed for the computation of
four-point scattering amplitudes of scalar (tachyon) vertex
operators are not elementary and we only study the appropriate
singularities and interpret them in terms of OPE.
As mentioned in section \ref{HppPenrose} the Hpp-wave with
$\widehat{\mathcal
H}_6$ affine Heisenberg symmetry that emerges in the Penrose limit
of $AdS_3\times S^3$ should be combined with extra degrees of
freedom in order to represent a consistent background for the
bosonic string. Quite independently of the initial values of
$k_{SL(2,\mathbb{R})}=k_1$ and $ k_{SU(2)}=k_2$, one needs to combine the
resulting CFT that has $c=6$ with some internal CFT with
$c=20$. For definiteness let us suppose this internal CFT to
correspond to flat space ${\bf R}^{20}$ or to a torus $T^{20}$, but
this choice is by no means crucial in the following.

In a covariant approach, such as the one followed throughout the
paper, string states correspond to BRS invariant vertex operators.
As usual, negative norm states correspond to unphysical
`polarizations'. These are absent for the scalar (tachyon) vertex
operators we have constructed in section \ref{spectrum}.
Let us focus on the left-movers. Starting from a
`standard' HW ($ \mu_\alpha p <1$ for $\alpha=1,2$) primary state
$|\Psi\rangle$ of $\widehat{\mathcal H}_6$, the Virasoro
constraints \be L_n|\Psi\rangle = 0 \ , \hspace{2cm} {\rm for} \quad n>0 \ ,
\ee
together with \be L_0|\Psi\rangle =|\Psi\rangle \ , \ee project the
Hilbert space on positive norm states. The mass-shell condition
becomes \be h^a_{p,\hj} + h_{int} + N =1 \ , \label{dimcon} \ee where
$N$ is the total level, $h_{int}$ is
the contribution of
the internal CFT, \ie \ $h_{int} = |\vec{p}|^2/2$ and
for $p\neq 0$
\be h^{\pm}_{p,\hj} = \mp p \hj + {1\over 2} \sum_{\alpha=1}^2
\mu_\alpha p (1 - \mu_\alpha p) \ , \ee while for $p=0$,  \be
h^{0}_{s,\hj} = {1\over 2} s^2 = {1\over 2} \sum_{\alpha=1}^2
s_\alpha^2 \ . \ee

Outside the range $\mu_\alpha p <1$ one has to
consider spectral flowed representations when $\mu_1=\mu_2=\mu$ or
MHW representations when $\mu_1\neq \mu_2$, as discussed in section
\ref{spectrum}.
Let us concentrate for simplicity on $\mu_1=\mu_2=\mu$ with
enhanced (non-chiral) $SU(2)_I$ invariance. In this particular
case, spectral flow yields states with \be h^{\pm, w}_{p,\hj} =
\mp \left ( p + {w \over \mu} \right ) \hj + \mu p (1 - \mu p) \mp w \lambda
\ ,
\ee
where $\lambda = n_- - n_+$ is the total `helicity' and, for
$p=0$, \be h^{0, w}_{s,\hj} ={w \over \mu} \hj - {1\over 2} s^2 -
w \lambda \ . \ee
The physics is similar to the case of the NW background
\cite{dak}: whenever $\mu p$ reaches an integer value in string
units, stringy effects become important and one has to resort to
spectral flow in order to make sense of the resulting state
\cite{hik}. The string feels no confining potential and is free to
move along the `magnetized planes'. The analysis of $AdS_3$ leads
qualitatively to the same conclusions \cite{mo1}. Spectral flowed
states can appear both in intermediate channels and as external
legs. Even though in this paper we have only considered  correlation
functions with states in
highest weight representations with $\mu|p|<1$ as external legs, it is not
difficult to generalize our results to include spectral
flowed external states along the lines of \cite{dak}.

In order to compute covariant string amplitudes in the Hpp-wave
one has to combine the correlators computed in sections
\ref{3point}, \ref{4point} and \ref{wakimoto} with the
contributions of the internal CFT and of the bosonic $b,c$
ghosts. Contrary to the AdS case discussed in \cite{mo1,mo3}, we
do not expect any non-trivial reflection coefficient in the
Hpp-wave limit, so, given the well known normalization problems in
the definition of two-point amplitudes, let us start considering
three-point amplitudes.  As it was shown in section 4.1,
three-point functions in the Hpp-wave precisely agree with those
resulting from the Penrose limit of three-point functions in
$AdS_3\times S^3$.

The irreducible three-point coupling can be directly extracted
from the tree-point correlation functions computed in section
\ref{3point}. Trading the integrations over the insertion points
for the volume of the $SL(2,\mathbb{C})$ global isometry group of the
sphere and combining with the trilinear coupling $T_{IJK}(h_i)$ in
the internal CFT one simply gets \be {\cal A}_{abc}^{IJK}(\nu_i,
x_i; h_i) = K_{abc}(\nu_i, x_i) C_{abc}(\nu_i) T_{IJK}(h_i) \ ,
\ee where $a_i = \pm,0$, $\nu_i$ denote the relevant quantum
numbers and the $\delta$-functions associated to the conservation
laws are understood. Except for $T_{IJK}(h_i)$ all the relevant
pieces of information can be found in section 4. For ${\mathcal
M}= {\bf R}^{20}$ or $T^{20}$, $T_{IJK}(h_i)$ is essentially
purely kinematical, \ie \ $\delta(\sum_i \vec{p}_i)$. Other
consistent choices require a case by case analysis. Depending on
the kinematics, amputated three-point amplitudes can be
interpreted as decay or absorption rates. In particular
kinematical regimes (for the charge variables) they allow one to
compute mixings, to determine the $1/k \approx g_s$ corrections to the
string spectrum in the Hpp-wave and to address the problem of
identifying `renormalized' BMN operators
\cite{bmn,ppreviews,opmix}.

Additional insights can be gained from the study of four-point
amplitudes. In particular the structure of their singularities
provides interesting information on the spectrum and couplings of
states that are kinematically allowed to flow in the intermediate
channels. Needless to say, one would have been forced to discover
spectral flowed states or non-highest weight states even if one
had not introduced them in the external legs.

As usual, $SL(2,\mathbb{C})$ invariance allows one to fix three of the
insertion points and integrate over the remaining one or rather
their $SL(2,\mathbb{C})$ invariant cross ratio denoted by $z$ in previous
sections. Schematically
\be {\cal A}_4 = \int d^2z |z|^{\sigma_{12} - 4/3}
|1- z|^{\sigma_{14} - 4/3} K(x_i, \nu_i) G_{Hpp}(\nu_i,x_i, z)
G_{\mathcal M}(h_i,z) \ , \ee where, for a flat ${\mathcal M}$,
$\sigma_{ij} = \kappa_{ij} + \vec{p}_i\cdot \vec{p}_j$ with
$\kappa_{ij}$ defined in section 4.

At present, closed form expressions for ${\cal A}_4$ are not
available. Still the OPE allows one to extract interesting physical
information. Let us consider, for a flat ${\mathcal M}$, the two
cases ${\cal A}_{+++-}$ and ${\cal A}_{+-+-}$. The relevant $\hat {\cal H}_6$
four-point functions have been computed both solving the KZ equation (in
section \ref{4point}) and by means of the Wakimoto representation
(in section \ref{wakimoto}).
Expanding ${\cal A}_{+++-}$ in the s-channel yields \bea &&{\cal
A}_{+++-} = \int d^2z |z|^{2(h_{12}-2)} \sum_{q=0}^Q
C_{++}{}^+(\nu_1,\nu_2; q) C_{+-}{}^-(\nu_3,\nu_4; Q-q) \nonumber \\
&&|z|^{-2q(p_1+p_2)} ||x_{12}||^{2q} ||x_{13}||^{2(Q-q)} + ...
\eea
where $q= q_1$, $h_{12}= h^+(p_1 + p_2, \hj_1 +\hj_2) + {1\over
2} (\vec{p}_1 + \vec{p}_2)^2$. Studying the $z$ integration near
the origin determines the presence of singularities whenever
$h_{12} - q(p_1+p_2) = 1 - N$ that coincides with the mass-shell
condition for the intermediate state in the $V^+$
representation.
The amplitudes ${\cal A}_{+-+-}$ are more interesting in that they
feature the presence of logarithmic singularities in the s-channel
when $p_1 = p_2$ and $p_3 = p_4$. The amplitude factorizes in the
continuum of type 0 representations parameterized by $s$ \be {\cal
A}_{+-+-} = \int d^2z |z|^{2(h_{12}-2)} \int s^3 ds
C_{+-}{}^0(p_1,s) C_{+-}{}^0(p_3,s) |z|^{s^2} (||x_{13}||||x_{24}||)^{2k} I_{|k|}+
... \ , \ee
where in the present case $h_{12}={1\over 2} (\vec{p}_1 +
\vec{p}_2)^2$.
Using the explicit form of the OPE coefficients determined in
section 4, and integrating $z$ in a small disk around the origin
yields \be {\cal A}_{+-+-} \approx \int d^2z |z|^{2h_{12}-4-2L}
\Psi(p_1,p_3)^{L+1}|\exp{(p_3 x z + x \Psi(p_1,p_3))} \sum_{q=0}^{\infty}
{|x \Psi(p_1,p_3)|^{2q} \over {q! (Q + q)!}} \ , \ee
where as usual $Q = L/\mu =-\sum_i \hj_i/\mu$ and $\Psi(p_1,p_3)
= [- \log|z|^2 - 4\psi(1) - \psi(p_1) - \psi(1-p_1)-
\psi(p_3) - \psi(1-p_3)]^{-1}$. For $q=Q=0$ one has \be {\cal
A}_{+-+-} \approx \int_{|z|<\epsilon} {d|z| \over |z|^\delta
\log|z|} \ , \ee where $\delta = 3 - 2h_{12}$ that converges for
$\delta<1$ but diverges logarithmically as ${\cal A}_{+-+-}
\approx \log(h_{12} -1)$ for $\delta\approx 1$. The logarithmic
branch cut departing from $h_{12}=1$ signals the presence of a
continuum mass spectrum of intermediate states with $s=0$.
Expanding in the u-channel for $p_1 +p_3=w$ one can proceed
roughly in the same way and identify the continuum of intermediate
states in spectral flowed type 0 representations. They signal the
presence of branch cuts for each string level.

\renewcommand{\theequation}{\arabic{section}.\arabic{equation}}
\section{The holographic correspondence}
\label{holography}

Having explicit control on the detailed action of the Penrose limit on
string theory in $AdS_3\times S^3$, we can employ the original
$AdS_3/CFT_2$ recipe to provide a concrete formula for the
holographic correspondence in the Hpp-wave background. On the
string side we end up with S-matrix elements as anticipated
earlier \cite{kirpio} and defined unambiguously in \cite{dak}. On
the $CFT_2$ side we can produce an explicit formula for
the Penrose limit of CFT correlators, to be compared with the
string theory S-matrix elements.

The key ingredients of such a holographic formula are:

$\bullet$ The original $AdS_3/CFT_2$ equality between ``S-matrix"
elements\footnote{These are not the standard S-matrix elements,
but their closest analogue in AdS. They can be defined as the
on-shell action evaluated on a solution of the (quantum) equations
of motion with specified sources on the boundary. For $AdS_3$ such
elements were conjectured by Maldacena and Ooguri \cite{mo3}.} for
vertex operators in Minkowskian signature $AdS_3$ and
CFT correlation functions. Introducing two charge variables $\vec x$
for $SL(2,\mathbb{R})$ and as many $\vec y$ for $SU(2)$, the ``S-matrix elements"
depend on both $\vec x$  and $\vec y$. On the CFT side, $\vec x$ represent
the positions of CFT operators ${\cal O}_{l,\tl}(\vec x,\vec y)$, while
$\vec y$ are charge variables for the $SU(2)_L\times SU(2)_R$ R-symmetry.
The conformal weight of the operators ${\cal O}_{l,\tl}$ is given by $\Delta=l$.

$\bullet$ The limiting formulae (\ref{pp+}), (\ref{pp-}) and
(\ref{pp0}) that describe the
precise
way operators of the original theory map to the operators of the
pp-wave theory under the Penrose contraction.

In the expressions below, $\vec
z_i$ are the coordinates of the vertex operators on the string world-sheet,
$\vec x_i$ are the $SL(2,\mathbb{R})$ charge variables, that represent the insertion
points on the boundary, and $\vec y_i$ are the $SU(2)$
R-charge variables. $\Psi^{\pm}_{l}\left(\vec z,\vec x\right)$ are
$SL(2,\mathbb{R})$ primary fields of string theory on $AdS_3$ corresponding to the
${\cal D}^{\pm}_l$ representations, $\Omega_{\tilde l}\left(\vec
z,\vec x\right)$ are $SU(2)$ primary fields of string theory on $S^3$
corresponding to the $SU(2)$ representation of spin $\tilde l$,
and $\Psi^{0}_{l,\a}\left(\vec z,\vec x\right)$ are the
$SL(2,\mathbb{R})$ primary fields of string theory corresponding to the
continuous representations of spin $l$. We neglect the internal
CFT part of the operators as it is not relevant for the structure
of our formulae.

The left and right charge variables $x,\bar x$ are related to the Cartesian
ones used here
by $x=x^1+ix^2, \bar x=x^1-ix^2$.
Thus, the transformation that inverts the chiral charge variables, $x\to
1/x, \bar x\to 1/\bar x$
corresponds in the cartesian basis to $\vec x \to \vec x^c/|\vec x|^2$ where
the superscript stands for a parity transformation,
$(x^1,x^2)^c=(x^1,-x^2)$. Since we consider lorentzian $AdS_3$
also a Minkowski continuation of the charge variables is necessary,
and this can readily be implemented
in the CFT correlators by $x\to x^+,\bar x\to x^-$.

We will denote by ${\cal O}_{l,\tilde l}(\vec x,\vec y)$ operators in the
CFT that correspond to the appropriate ones in  $AdS_3$
\be \Psi_{l}\left(\vec z,\vec x\right)\Omega_{\tilde l}\left(\vec
z,\vec y\right) \Leftrightarrow {\cal O}_{l,\tilde l}\left({\vec x,\vec
y}\right) \ . \ee
The $AdS_3$ ``S-matrix elements" are functions of the spins $(l,\tl)$
as well as of the charge variables $\vec x_i,\vec y_i$.
They can be obtained by standard techniques by integrating the CFT
correlators appropriately over the positions of the vertex operators
\cite{mo3}.
We will split the $AdS_3$ states into three families, distinguished
by the type of ${\cal H}_6$ representation they will asymptote to
in the Penrose limit,
namely $\Phi^+$, $\Phi^-$ and $\Phi^0$.
Thus the starting string ``S-matrix elements" are of the form \be
S^{AdS_3}_{N_{\pm,0}}(l_i,\tilde l_i,\vec x_i,\vec
y_i|l_j,\tilde l_j,\vec x_j,\vec y_j|l_k,\a_k,\tilde l_k,\vec
x_k,\vec y_k) \ , \ee
where the index $i = 1, ..., N_+$ labels the operators that asymptote
to the $\Phi^+_{p_i,\jh_i}$ operators,  the index $j = 1, ..., N_-$ labels the
operators that asymptote
to the $\Phi^-_{p_j,\jh_j}$
operators and the index $k = 1, ..., N_0$ labels the operators that asymptote
to the $\Phi^0_{s^1_k,s^2_k,\jh_k}$ operators.
As shown in section (\ref{3point}), by taking the Penrose limit the
$AdS_3 \times S^3$ S-matrix elements asymptote to the pp-wave S-matrix elements
we computed as
$$ \lim_{k_1\to\infty\atop k_2\to \infty} \prod_{i=1}^{N_+}
\left( {k_1\over |\vec x_i|^2}\right)^{-2l_i}~ \left( {k_2\over
|\vec y_i|^2}\right)^{2\tilde l_i}
\prod_{k=1}^{N_0} |\vec x_k|^{-2l_k}|\vec y_k|^{2\tl_k}
\times
$$
\be\times
S^{AdS_3}_{N_{\pm,0}} \left( l_i,\tilde l_i,{\sqrt{k_1}\vec
x_i^c\over  | \vec x_i|^2},{\sqrt{k_2}\vec y_i^c\over  |\vec
y_i|^2} \left | l_j,\tilde l_j,{\vec x_j\over \sqrt{k_1}},{\vec
y_j\over \sqrt{k_2}} \right | l_k,\a_k,\tilde l_k,\vec x_k,\vec
y_k \right )= \label{sads3} \ee
$$
=C_{N_+,N_-,N_0}(k_1,k_2)~
S^{Hpp}_{N_{\pm,0}}(p_i,\jh_i,\vec x_i,\vec y_i| p_j,\jh_j,\vec x_j,\vec
y_j|s_{k}^{1,2},\jh_k, \vec x_k,\vec y_k) \ . $$
In the previous formula the limit on the spins is taken as
explained in section \ref{3point}.
For the first two classes of operators (labeled by $i$ and $j$) we have
\be
l = \frac{k_1}{2} \m_1 p - a \ , \hspace{1cm} \tl = \frac{k_2}{2} \m_2 p - b \ ,
\ee
with the subleading terms $a$ and $b$ related to $\jh$ in the limit as follows
\be
\jh_i = -\m_1 a_i + \m_2 b_i \ , \hspace{1cm} \jh_j = \m_1 a_j - \m_2 b_j \ .
\ee
For the third class of operators we set
\be
l = \frac{1}{2} + i \sqrt{\frac{k_1}{2}} s_1 \ , \hspace{1cm}
\tl =  \sqrt{\frac{k_2}{2}} s_2 \ ,
\ee
and in the limit  $\jh_k$ is given by the fractional part of
the $SL(2,\mathbb{R})$ spin $\jh_k = - \m_1 \a_k$.
The coefficients $C_{N_+,N_-,N_0}(k_1,k_2)$ are divergent in the limit $k_{1,2}\to
\infty$ and can be computed in principle directly. Using the results obtained
in section \ref{3point} we have for instance
\be
C_{2,1,0}(k_1,k_2) =  \sqrt{k_1 k_2} \ .
\ee
By employing the holographic recipe of AdS/CFT we can now write the relation
between pp-wave S-matrix elements and limits
of CFT correlators\footnote{We ignore type $V^0$ operators since, although
their definition and dynamics are clear on the string theory side, they are
less clear in the CFT side.
They are  related to the continuous spectrum and the associated
instabilities of the $NS5/F1$ system in analogy with
the discussion in \cite{sw}.}
\be S^{Hpp}_{N_{\pm}}(p_i,\jh_i,\vec x_i,\vec y_i| p_j,\jh_j,
\vec x_j,\vec y_j)=\lim_{k_1\to\infty\atop k_2\to \infty} {\prod_{i=1}^{N_+}
\left( {k_1\over |\vec x_i|^2}\right)^{-2l_i}\prod_{j=1}^{N_-}\left( {k_2\over
|\vec y_j|^2}\right)^{2\tilde l_j}\over C_{N_+,N_-}(k_1,k_2)}\times \ee
$$
\left \langle \prod_{i=1}^{N_+}
{\cal O}_{l_i,\tilde l_i}\left(\sqrt{k_1}{\vec x_i^c\over |\vec
x_i|^2},\sqrt{k_2}{\vec y_i^c\over |\vec
y_i|^2}\right)\prod_{j=1}^{N_-} {\cal O}_{l_j,\tilde l_j}\left({\vec
x_j\over \sqrt{k_1}},{\vec y_j\over \sqrt{k_2}}\right)
\right \rangle \ .
$$
The $SL(2,\mathbb{R})$ spin is the conformal dimension of the CFT operator while the
$SU(2)$ spin determines its transformation properties under the $SU(2)$ R-symmetry.
The level $k$ in the space-time CFT is interpreted as the number of $NS5$ branes used
to build the background \cite{bort}.

The interpretation of the limit in the CFT is as follows.
CFT operators that asymptote to $V^-$ representations (with negative values
of $p^+$)
have their position and charge variables scaled to zero.
Operators that asymptote to $V^+$ representations (with positive values of
$p^+$)
are instead placed at antipodal points and then their positions are scaled
to infinity.
Finally all the spins are scaled as indicated and there is an  overall
renormalization.
The limit of the two-point functions of the CFT
is particularly simple. In this case $C_{1,1,0}=1$ and we obtain in the Penrose limit
\be
S(p_1,\jh_1,\vec x_1,\vec y_1|p_2,\jh_2,\vec x_2,\vec y_2)=
\exp\left[-\mu_2p(y_1y_2+\bar y_1\bar y_2)-
\mu_1p(x^+_1x^+_2+x^-_1x^-_2)\right] \ ,
\ee
where $\vec y_i$ are in Euclidean space and $\vec x_i$ are in Minkowski space .

\renewcommand{\theequation}{\arabic{section}.\arabic{equation}}
\section{Outlook}
\label{outlook}

In this paper we have computed tree-level (sphere) bosonic
string amplitudes in the Hpp-wave limit of $AdS_3\times S^3 \times
{\mathcal M}_{20}$ supported by NS-NS 3-form flux. For simplicity,
we have only considered scalar `tachyon' vertex operators with no
excitation in the internal world-sheet CFT describing ${\mathcal
M}_{20}$. The present results generalize the bosonic string
amplitudes obtained for the NW model that arises in the Penrose
limit of the near horizon geometry of a stack of penta-branes
\cite{dak}. Preliminary results for the simplest `extremal'
amplitudes of the type $\langle +++ -\rangle$ have been presented
in \cite{oz}. We have heavily relied on current algebra techniques
on the world-sheet and confirmed for the present case, with affine
$\widehat{\mathcal H}_6$ Heisenberg symmetry, the agreement with
the free-field Wakimoto realization found in \cite{cfs} for the NW
model, enjoying an affine $\widehat{\mathcal H}_4$  symmetry.

We have discussed both the $SU(2)$ symmetric case ($\mu_1 =
\mu_2$) and the general case ($\mu_1 \neq \mu_2$) and observed
that the corresponding exactly marginal deformations interpolate
between the generic 6-d Hpp-wave ($\mu_1 \neq \mu_2$), the
(super)symmetric one ($\mu_1 = \mu_2$) and the NW model ($\mu_1 =
0$ or $\mu_2=0$) or even flat space-time ($\mu_1 = 0$ and
$\mu_2=0$) very much as the `null deformation' discussed in
\cite{ikp} interpolates between $AdS_3\times S^3$ and $R^+\times
S^3$ with a linear dilaton before any Penrose limit is taken. The
space-time counterpart of the world-sheet RG flow is the
condensation / evaporation of fundamental strings \cite{ikp}.
We have derived covariant
bosonic string amplitudes on the sphere and shown that they are
well defined even for $p^+=0$ states, which are difficult if not
impossible to analyze in the light-cone gauge. String amplitudes
expose singularities that admit a sensible physical interpretation
in terms of OPE, very much as in the closely related NW model
\cite{dak}, and precisely match the ones resulting from the
`Saletan contraction' \cite{sfetsos} $k_1,k_2\to \infty$ with
$\mu_1^2 k_1 = \mu_2^2 k_2$ of $SL(2,\mathbb{R})_{k_1}\times SU(2)_{k_2}$.
We have thus provided further evidence for the consistency of the
BMN limit \cite{bmn} in this setting. A crucial role has been
played by the complex charge variables that can be introduced for
any group in order to compactly encode the content of (in)finite
dimensional irreps \cite{zf,tesch,arutsoka,dak}. While for the
$SL(2,\mathbb{R})_L\times
SL(2,\mathbb{R})_R$ algebra, underlying $AdS_3$, $x$ and $\bar{x}$ can be viewed as
coordinates on the 2-d boundary, in the case of the ${\mathcal
H}_{2+2n}$ algebra, underlying a pp-wave, $x^\alpha$ and $x_\alpha$ become
coordinates on a $2n$-dimensional `holographic' screen
\cite{kirpio} that replaces the one-dimensional null boundary
representing the `true' geometric boundary in the Penrose limit
\cite{bn,dorn1}.

At any finite but large value of $k$ (\ie \ the radius or any other
contraction parameter) one finds (or rather expects) a perfect
matching between string amplitudes in $AdS_3\times S^3 \times
{\mathcal M_{20}}$ and correlation functions in some boundary
CFT$_2$. Only when the contraction is fully performed target space
conformal invariance should be replaced with the relevant
${\mathcal H}_{2+2n}$ Heisenberg symmetry, along with its `accidental' external
automorphisms. For the bosonic string, there is no obvious
candidate for a `dual' boundary CFT$_2$. The naive guess would be
a $\sigma$-model on a resolution of $Sym_N
({\mathcal M_{20}})$ with $N= N_1 N_{21}$ with $N_1$ the number of
fundamental strings wrapped around an $S^1$ and smeared in
$M_{20}$ and $N_{21}$ the 21-branes wrapped around $S^1\times
M_{20}$. Anyway, despite the presence of tachyons and other
limitations, for states with large R-charge and corresponding to
the identity sector of the world-sheet CFT describing $M_{20}$,
tree-level (sphere) amplitudes should display the relevant
pattern: conformal invariance $\rightarrow$ Saletan contraction $\rightarrow$ Heisenberg
symmetry. Indeed our results at least qualitatively point in this
direction.

In order to be more quantitative one should consider the (type
IIB) superstring where the candidate dual boundary CFT  is the
${\cal N}=(4,4)$ $\sigma$-model on the (hyperk\"ahler resolution
of) $Sym_N({\mathcal M_{4}})$
with $N= N_1 N_{5}$ with $N_1$ the number
of fundamental strings wrapped around an $S^1$ and smeared in
$M_{4}$ and $N_5$ the number of NS5 branes wrapped around $S^1\times
M_{4}$ \cite{malda}.  Despite some initial success for the
matching of the KK supergravity spectrum with the spectrum of
chiral primary operators \cite{deboer}, a stringy exclusion
principle \cite{maldastrom}, which is related to the existence of
a maximal allowed R-symmetry charge, even for multi-particle
states (differently from the more familiar CFT$_4$ case!) has
stimulated some reconsideration. In particular, it is widely
believed that the `symmetric orbifold point' of the boundary
CFT$_2$, that should be the analogue of the `higher spin
enhancement point' $g_{YM}=0$ of ${\cal N}=4$ SYM in $D=4$
\cite{bosund,witths,ss}, does not coincide with the locus in the
moduli space where the string description is under control, at
least in the case of NS-NS flux \cite{gks,kutseib}. The latter
should in fact correspond to a singular CFT$_2$ due to the
presence of non-compact directions in the target space of the
$\sigma$-model related to the possibility of string emission from
penta-branes \cite{sw}, as mentioned above. Indeed the presence of
a continuous spectrum of long strings in $AdS_3$ and their
images under spectral flow seems to point in this direction
\cite{mo1,mojs2,mo3}. In particular some of the missing chiral
primaries \cite{kll,ags}, usually associated to short strings,
may have reached the unitary bound, re-combined with other states
with the proper quantum numbers and disappeared in the continuum
as long strings. Indeed, following the analysis of
\cite{bms,bbms}, \ie \ extrapolating the string spectrum to the
relevant `enhancement / orbifold' point, it has been recently
argued that this generalized Higgs mechanism is at work
\cite{son}. Considerations of the dynamics in pp-wave backgrounds
\cite{gavnar,gms,lunmat,beis} have certainly helped pursuing this
line of thought. Once again, long strings are associated with
states with $p^+ \in \mathbb{Z}$ which require a covariant description,
being related by spectral flow to the $p^+ = 0$ representations.

Alternatively, one may consider turning on R-R fluxes which should
effectively compactify the target space \cite{sw,mo1,mo3}. The
hybrid formalism of Berkovits, Vafa and Witten \cite{bvw} seems
particularly suited to this purpose as it allows the computation
of string amplitudes, at least for the massless modes
\cite{bobdol}, and the study of the Penrose limit in a covariant
way \cite{berkpp}. The pure spinor formalism \cite{berkovits}
might be needed for ${\mathcal M}_4=T^4$ due to the enhanced susy
(16 $\to$ 24). The mismatch for 3-point functions of chiral
primaries (or rather their superpartners)\footnote{This may
actually be due to the  reduced number of susy's (16 instead of 32
!) and the consequent lack of a non-renormalization theorem for
these couplings.} \cite{AFP1,AFP2} calls for additional
investigation in this direction and a careful comparison with the
boundary CFT results of \cite{lunmat}. Once again, the BMN limit
\cite{bmn} may shed some light on this issue as well as on the
short-distance logarithmic behavior found in \cite{mo3} for AdS
and in \cite{dak} for NW that require a resolution of the operator
mixing along the lines of \cite{opmix} or a scattering matrix
interpretation \cite{freedgur}.

It would thus be very interesting and important to extend the
present analysis to the superstring, compute scattering amplitudes
in the Hpp-wave and study their (super)symmetry properties.
In principle, one would like to address
some of the above issues (spectrum, trilinear couplings and
operator mixing) in a more quantitative way and possibly
reformulate the holographic duality in the Penrose limit directly
in terms of propagators along the lines of \cite{lunmat,dorn2}
that should further clarify the role of the charge variables as
coordinates in a holographic screen \cite{bdkz2}.

In summary, we would like to argue that the BMN limit of physically
sensible correlation functions is well defined and perfectly consistent,
at least for the CFT dual to $AdS_3\times S^3$. In particular it
should not lead to any of the difficulties encountered in the case
of ${\cal N}=4$ SYM as a result of the use of perturbative schemes
or of the light-cone gauge. The case of
the Hpp-wave supported by NS-NS fluxes being under control at each
step (before and after the Penrose limit is taken) could prove to
be a source of extremely useful insights in holography and the
duality between string theories  and field theories.

\vskip 2cm

\acknowledgments The authors are grateful to O. Aharony, I.~Bakas,
N.~Berkovits, N.~Constable, A. Giveon, C.~Hull, J.~F.~Morales, K.~Panigrahi, I.~Plefka,
E.~Sezgin, K.~Sfetsos, E. Sokatchev, J. Sonnenschein, Ya.~Stanev for discussions.

This work was partially supported by EU contracts
HPRN-CT-2000-00122, HPRN-CT-2000-00148, HPRN-CT-2000-00131,
MIUR-COFIN contracts 2001-025492, 2003-023852, INFN, INTAS
contract N~2000-254, NSF grant PHY99-07949, and the NATO contract
PST.CLG.978785. The work of G.~D. is supported by a European
Commission Marie Curie Postdoctoral Fellowship, contract HMPF CT
2002-01908. The work of O.~Z. was also supported by a ICSC-World
Laboratory fellowship.

E.~K. would like to thank KITP for hospitality during part of this
work. M.~B. and O.~Z. would like to thank LPTHE at Jussieu for hospitality
during completion of this work.

\newpage
\centerline{\bf APPENDIX}
\appendix
\renewcommand{\theequation}{A.\arabic{equation}}
\section{ The $\sigma$-model view point}
\label{app A}


{}Elements of the ${\bf H}_6$ Heisenberg group can be parametrized as \cite{cfs}
  \beq
g(u,v,\gamma^{\alpha},\bar \gamma_{\alpha})=e^{\frac{\gamma^{\alpha}}{\sqrt
2}P^+_{\alpha}}e^{uJ-vK}e^{\frac{\bar \gamma_{\alpha}}{\sqrt
2}P^{-\alpha}}\ .
  \eeq
As usual the $\sigma$-model action can be written in terms of the Maurer-Cartan forms and reads
\beq \label{wakact1} S = \frac{1}{2 \pi}\,\int d^2\sigma \left( -\pd u
\bar{\pd} v + \sum_{\a=1}^2 e^{i\mu_\a u} \pd\bar\gamma_\a
\bar\pd\gamma^\a \right), \eeq
where we have used $\langle J,K\rangle=1$ and $\langle
P^{+}_{\alpha},P^{-\alpha}\rangle=2$.
The metric and $B$ field are then given by
{\setlength\arraycolsep{2pt} \beqa
ds^2&=&-2dudv+2\sum_\a e^{i\mu_\a u} d\gamma^{\alpha}d\bar{\gamma}_{\alpha}\ ,\\
B&=&-du\wedge dv+\sum_\a e^{i\mu_\a u}d\gamma^{\alpha}\wedge
d\bar{\gamma}_{\alpha}\ . \eeqa}

Two auxiliary fields $\beta_{\alpha}$ and
$\bar{\beta}^{\alpha}$, defined by the OPE's
  \beq
\beta_{\alpha}(z)\gamma^{\beta}(w)\sim\frac{\delta_{\alpha}^{\beta}}{z-w}
\ . \eeq
complete the ghost-like systems that appear in the Wakimoto
representation.

With the help of $\beta_{\alpha}$ and
$\bar{\beta}^{\alpha}$, the action can be written as
  \beq \label{wakact2} S =
\frac{1}{2\pi} \int \, d^2 z \left( -\pd u \bar{\pd} v + \sum_{\a=1}^2 [
\bar\beta^{\alpha}  \pd \bar\gamma_{\alpha} + \beta_{\alpha}
\bar{\pd} \gamma^{\alpha} - e^{-i\mu_\a u} \beta_{\alpha}
\bar\beta^{\alpha} ] \right)\ , \eeq
that gives us back (\ref{wakact1}) upon using the equations of motion for $\beta_{\alpha}$ and
$\bar{\beta}^{\alpha}$.

In the Wakimoto representation, the currents can be written  as \cite{cfs}
{\setlength\arraycolsep{2pt}\beq \begin{array}{rcl}
P^+_{\alpha}(z) &=&
-\beta_{\alpha}(z), \\
P^{-\alpha}(z)&=& -2(\partial \gamma^{\alpha} + i
\partial u \gamma^{\alpha})(z), \\
J(z) &=& -(\partial v - i \sum_\a \mu_a \beta_{\alpha} \gamma^{\alpha})(z),  \\
K(z) &=& - \partial u(z)\ ,
\end{array}\eeq}
in agreement with the result of section \ref{wakimoto}.
A simple identification of the $\bf H_6$ group parameters and
the string coordinates, recast the metric in the more standard form
of (\ref{pp2trans}).
Generalizing the results of \cite{cfs}, it is easy to show that string coordinates
and Wakimoto fields are related as follows
{\setlength\arraycolsep{2pt}\beq \begin{array}{rcl}
u(z,\bar{z}) &=& \displaystyle{ u(z) + \bar u(\bar{z})}, \\
v(z,\bz)  &=& v(z) + \bar
v(\bar{z})+2i\bar{\gamma}_{L\alpha}(z) \gamma_R^{\alpha}(\bar{z}),\\
w^{\alpha}(z,\bz) &=&\displaystyle{ e^{-{i}\mu_\a u(z)}
[e^{i\mu_\a u(z)}\gamma_L^{\alpha}(z)
+ \gamma_R^{\alpha}(\bar{z})] },\\
\bar{w}_{\alpha}(z,\bar{z})&=&\displaystyle{e^{+{i}\mu_\a u(z)}[\bar{\gamma}_{L\alpha}(z)
+e^{i\mu_\a\bar u(\bar{z})} \bar{\gamma}_{R\alpha}(\bar{z})]} \ .
\end{array}\eeq}

\end{document}